\documentclass[paper]{JHEP}
\usepackage[centertags]{amsmath}
\usepackage{latexsym,amssymb}
\usepackage{graphicx}
\usepackage{amsfonts} \usepackage{amssymb} \usepackage{amsthm}
\usepackage{psfrag}

\newcommand{\eq}[1]{Eq.~(\ref{#1})}

\newcommand{\lvev}{\Big\langle\hskip -5pt\Big\langle}
\newcommand{\rvev}{\Big\rangle\hskip -5pt\Big\rangle}
\newcommand{\vev}[1]{\langle #1 \rangle}

\newcommand{\dslash}{{\slash \kern -1ex \partial}}

\def\beq{\begin{equation}}
\def\eeq{\end{equation}}
\newcommand{\uno}{\mbox{1\!\negmedspace1}}
\newcommand{\beqa}{\begin{eqnarray}}
\newcommand{\eeqa}{\end{eqnarray}}
\newcommand{\bea}{\begin{eqnarray}}
\newcommand{\eea}{\end{eqnarray}}

\def\ii{{\rm i}}
\def\ee{\mathrm{e}}
\def\comm#1#2{\left[ #1, #2\right]}

\def\Tr{{\rm Tr}}

\title{Instanton Calculus In R-R Background And The Topological String
}

\author{
Marco Bill\'o,
Marialuisa Frau\\
Dipartimento di Fisica Teorica, Universit\`a di Torino\\
Istituto Nazionale di Fisica Nucleare - sezione di Torino \\
via P. Giuria 1, I-10125 Torino
}
\author{
Francesco Fucito\\
Istituto Nazionale di Fisica Nucleare - sezione di Roma 2\\
Dipartimento di Fisica, Universit\`a di Roma Tor Vergata\\
Via della Ricerca Scientifica, I-00133 Roma, Italy
}
\author{
Alberto Lerda\\
Dipartimento di Scienze e Tecnologie Avanzate \\
Universit\`a del Piemonte Orientale,
I-15100 Alessandria, Italy\\
Istituto Nazionale di Fisica Nucleare  - sezione di Torino,
Italy
}
\abstract{We study a system of fractional D3 and D(--1) branes
in a Ramond-Ramond closed string background and show that it describes
the gauge instantons of ${\cal N}=2$ super Yang-Mills theory
and their interactions with the graviphoton of $\mathcal{N}=2$ supergravity.
In particular, we analyze the instanton moduli space using string theory methods and
compute the prepotential of the effective gauge theory exploiting the localization methods
of the instanton calculus showing that this leads to the same information
given by the topological string. We also comment on the relation between our approach and
the so-called $\Omega$-background.}
\preprint{DFTT-07/2006\\ROM2F/2006/11}
\begin{document}
\setcounter{section}{0}
\section{Introduction}
The relationship between string theory and perturbative field theories has been
thoroughly investigated for many years.
The study of the non-perturbative effects
in string theory and their comparison with field theory
is instead much more recent.
In particular, only after the introduction of D branes it has been
possible to significantly improve our knowledge of the
non-perturbative aspects of string theory. From the open string
point of view, the D branes are hyper-surfaces
spanned by the string end points on which a supersymmetric gauge
theory is defined.
For instance, a stack of $N$ D3 branes in flat space supports
a four-dimensional ${\mathcal N}=4$ supersymmetric Yang-Mills gauge theory.
Adding a set of $k$ D(--1) branes  (also called D-instantons)
allows to describe instanton configurations of winding number
$k$ \cite{witten,douglas,Lawrence:1998ja,green,Green:2000ke}. In fact, the excitations of the
open strings stretching between two D-instantons or between
a D3 brane and a D-instanton, are in one-to-one
correspondence with the ADHM moduli
of the super Yang-Mills instantons, and their interactions
correctly account for the measure on the moduli space \cite{Dorey:2002ik}
\footnote{For an alternative string approach to gauge instantons based on tachyon condensation,
see for example Ref. \cite{Hashimoto:2005qh}}.

In a recent paper \cite{Billo:2002hm}, to substantiate the above remarks,
its has been shown how the computation of tree-level string
scattering amplitudes on disks with mixed boundary conditions for a
D3/D(--1) system leads, in the infinite tension limit
$\alpha^\prime\to 0$, to the effective action on the instanton moduli space
of the $\mathcal{N}=4$ supersymmetric Yang-Mills theory. Furthermore,
it has been proved that the very same disk diagrams also yield the
classical field profile of the instanton solution, and that these mixed
disks effectively act as sources
for the various components of the $\mathcal{N}=4$ gauge multiplet.
In this framework \cite{Fucito:2001ha,Dorey:2002ik}
it is also possible to describe theories with a
smaller number of supersymmetries and their instantons by suitably
orbifolding the D3/D(--1) system.
In particular, considering a configuration with fractional D3 branes and
fractional D-instantons in the orbifold
$\mathbb{R}^{4}\times \mathbb{C}\times\mathbb{C}^2/\mathbb{Z}_2$
one can describe a ${\mathcal N}=2$ super Yang-Mills theory in four dimensions
together with its instantons \cite{Lawrence:1998ja}. This is the system we study in this
paper.

Our aim is to generalize the construction mentioned above to encompass
the so-called multi-instanton equivariant calculus
\cite{Nekrasov:2002qd,Flume:2002az,Bruzzo:2002xf,Losev:2003py,Bruzzo:2003rw,Flume:2004rp,Nekrasov:2005wp}
and try to clarify some surprising properties that have been noticed in
the literature.
In the multi-instanton equivariant calculus, the instanton moduli action
is deformed by means of certain $\mathrm{U}(1)\times \mathrm{U}(1)$
transformations which act only on a subset of
the moduli and leave the ADHM constraints invariants
\footnote{In this paper we consider the case
in which the $\mathrm{U}(1)\times\mathrm{U}(1)$ transformations
are represented by $\ee^{+\ii\,\varepsilon}$ and
$\ee^{-\ii\varepsilon}$,
where $\varepsilon$ is the deformation parameter.}.
This deformation turns out to be the crucial
ingredient that allows the evaluation of the
instanton partition function $Z^{(k)}$ for arbitrary winding numbers $k$.
In fact, in the deformed theory the supersymmetry transformations on the
moduli space have only a finite number of isolated fixed points
so that it becomes possible to use the localization theorems
and compute exactly the integrals over the instanton moduli space.
The partition function obtained in this way turns out to be an even
function of the deformation parameter $\varepsilon$ and of the
vacuum expectation value (v.e.v.) $a$ of the chiral gauge
superfield of the $\mathcal{N}=2$ theory. Furthermore, by writing
\begin{equation}
Z(a;\varepsilon)= \sum_{k=1}^\infty Z^{(k)}(a;\varepsilon) =
\exp\left(\frac{\mathcal{F}_{\mathrm{n.p.}}
(a;\varepsilon)}{\varepsilon^2}\right)~~,
\label{zf}
\end{equation}
one finds that
\begin{equation}
\lim_{\varepsilon \to 0} \mathcal{F}_{\mathrm{n.p.}}(a;\varepsilon) =
\mathcal{F}(a)
\end{equation}
where $\mathcal{F}(a)$ is the non-perturbative Seiberg-Witten prepotential
\cite{Seiberg:1994rs}. At this point an obvious question arises:
what about the terms in $\mathcal{F}_{\mathrm{n.p.}}(a;\varepsilon)$
of higher order in $\varepsilon$ and their physical interpretation?
In Ref. \cite{Nekrasov:2002qd} (see also Refs.
\cite{Losev:2003py,Nekrasov:2005wp}) N. Nekrasov conjectured
that the terms of order $\varepsilon^{2h}$
describe gravitational corrections to the gauge prepotential
coming from closed string amplitudes on Riemann surfaces of genus $h$ and
that they should correspond
to F-term couplings in the $\mathcal{N}=2$ effective action of
the form
\begin{equation}
\int d^4x \,\,(R^+)^2  (\mathcal{F}^{+})^{2h-2}
\label{fh}
\end{equation}
where $R^+$ is the self-dual part of the
Riemann curvature tensor and $\mathcal{F}^+$ is the self-dual part of the
graviphoton field strength of $\mathcal{N}=2$ supergravity.
It is a well-known result of string theory that
the F-terms (\ref{fh}) are non-vanishing only on Riemann surfaces of genus $h$
and that they can be also computed with the topological string
\cite{Bershadsky:1993cx,Antoniadis:1993ze}. Using this information
Nekrasov's conjecture was tested to be true in Ref. \cite{Klemm:2002pa}
in the case of $\mathcal{N}=2$ super Yang-Mills with $\mathrm{SU}(2)$ gauge group.

In this paper we try to confirm and make more evident the above interpretation
by computing directly the couplings induced by a
(self-dual) graviphoton background on the instanton moduli space,
and by showing that they precisely match those that are induced
by the $\varepsilon$ deformation that fully localizes the instanton
integrals.
To do so we exploit the explicit string realization of the
$\mathcal{N}=2$ gauge theory and its instantons provided by a
D3/D(--1) system of fractional branes, and use the description
of the graviphoton of $\mathcal{N}=2$ supergravity
as a massless field the Ramond-Ramond closed string sector.
Then we determine how the graviphoton modifies the instanton
effective action by computing mixed open/closed string disk amplitudes.
We do this using the RNS formalism and the methods
already introduced in Ref. \cite{Billo:2004zq}
to study the non anti-commutative gauge theories \cite{Seiberg:2003yz}.
Even if it is a common belief that the RNS formalism is not suited to deal
with a R-R background, we show that this
is not completely true, and that a lot of information can actually
be extracted from this formalism in several cases, including the one
studied in this paper. In fact, by computing the instanton partition function in
a graviphoton background
with a constant self-dual field strength proportional to $\varepsilon$,
we can obtain through Eq. (\ref{zf}) a non-perturbative prepotential
$\mathcal{F}_{\mathrm{n.p.}}(a;\varepsilon)$ which coincides with the
one obtained with the multi-instanton equivariant calculus
\cite{Nekrasov:2002qd,Flume:2002az,Bruzzo:2002xf,Losev:2003py,Bruzzo:2003rw,Flume:2004rp}
but in which, by construction, the parameter $\varepsilon$ represents
the v.e.v. of the graviphoton field strength. Using standard superstring
methods \cite{Green:2000ke,Billo:2002hm}, we promote
$\varepsilon$ to a fully dynamical graviphoton field strength
$\mathcal{F}^+$ or even to the complete Weyl superfield
\cite{deWit:1984px} of which $\mathcal{F}^+$ is the lowest component.
Then, expanding the instanton-induced prepotential in powers
of this Weyl superfield, we obtain, among others, precisely the gravitational
F-terms of Eq. (\ref{fh}). It is worth noticing that in our approach
these terms arise from {disk} diagrams,
and specifically from $2h$ disks which, even if apparently
disconnected, must be effectively considered
as connected because of the
integration over the instanton moduli (see, for example,
the discussion in Section 6 of Ref. \cite{Billo:2002hm}).
Notice also that the Euler character of this topology
is the same as that of the world-sheet with $h$ handles
that is used in the topological string derivation of Eq.
(\ref{fh}).

In Refs. \cite{Losev:2003py,Nekrasov:2005wp} it has been argued
that the $\varepsilon$-deformation on the instanton moduli space
is due to a non-trivial metric, called $\Omega$-background, on the
gauge theory. At the linear order in the deformation this
$\Omega$-background is equivalent to a R-R
background.
In the present paper, we point out that, realizing the gauge theory and its
instantons via a fractional D3/D(--1) brane system, the parameter
$\varepsilon$  is directly related to the graviphoton.
In view of the above  considerations about the
gravitational F-terms (\ref{fh}) and of the connection with the
results of the topological string, we find this interpretation very natural.

This paper is organized as follows: in Section \ref{secn:N2}
we review how to derive the ${\mathcal{N}=2}$ action (both in the gauge and
in the instanton sectors) from tree-level open string scattering amplitudes
in a D3/D(--1) system, and discuss also how to
incorporate in the instanton effective action the v.e.v.'s of the scalar
gauge fields.
In Section \ref{secn:gravi} we analyze the instanton
moduli space in presence of a (constant) self-dual graviphoton
background by computing mixed open/closed string disk amplitudes,
and show that this R-R background induces precisely the same deformation
of the ADHM moduli space which fully localizes the instanton integrals.
In Section \ref{secn:omega} we compare our results with
those obtained with the deformed
ADHM construction, and show how to lift the graviphoton
deformation to the gauge theory action in four dimensions. We also
prove that the terms of this action that are linear in the graviphoton field
strength coincide with those produced by the $\Omega$-background
considered in Refs. \cite{Losev:2003py,Nekrasov:2005wp}.
Section \ref{secn:eff_action} is devoted to show
how the ${\cal N}=2$ effective action and the prepotential may be extracted
from the deformed instanton partition function, and how this compares with
the topological string approach.
Finally, in Section \ref{secn:concl} we present our conclusions, and
in Appendix A we list our notations and collect some formulas that are useful
for the explicit calculations.

\vskip 0.8cm
\section{${\cal N}=2$ gauge instantons from D3/D(--1) systems}
\label{secn:N2}

Instantons of charge $k$ in ${\cal N}=2$ theories with gauge group
$\mathrm{SU}(N)$ can be described within type IIB string theory
by considering systems of $N$
fractional D3 branes and $k$ fractional D(--1) branes at the fixed
point of the orbifold $\mathbb{R}^{4}\times \mathbb{C}\times
\mathbb{C}^2/\mathbb{Z}_2$. In this section we recall
this description, adapting to the
${\cal N}=2$ case the procedure discussed in
Refs. \cite{Billo:2002hm} and \cite{Billo:2004zq} for  ${\cal N}=4$
and ${\cal N}=1$ models.
Our notations and conventions, as well as the details of the $\mathbb{Z}_2$ orbifold
projection, are explained in Appendix \ref{secn:appA}.

\vskip 0.6cm
\subsection{The gauge sector}
\label{secn:gauge}

Let us consider type IIB string theory in
$\mathbb{R}^{4}\times \mathbb{C}\times
\mathbb{C}^2/\mathbb{Z}_2$ and place at the orbifold fixed point
a stack of $N$ fractional D3 branes that fill the four-dimensional
(Euclidean) space $\mathbb{R}^4$. The
massless excitations of open strings attached with both end points
to these branes describe the ${\cal N}=2$ gauge vector multiplet
in four dimensions, that comprises a gauge boson $A_\mu$,
two gauginos $\Lambda^{\alpha A}$ (with $\alpha,A=1,2$) and one complex scalar
$\phi$.
This field content can be assembled in a ${\cal N}=2$ chiral superfield
\begin{equation}
\Phi(x,\theta) = \phi(x) + \theta\Lambda(x) +\frac
12\,\theta\sigma^{\mu\nu} \theta
\,F_{\mu\nu}^+(x) + \cdots
\label{chiralsuperfield}
\end{equation}
where
\begin{equation}
\theta\Lambda(x) \equiv \theta^{\alpha A}\Lambda_{\alpha}
^{~B}(x)\epsilon_{AB}
~~~~,~~~~
\theta\sigma^{\mu\nu}\theta
\equiv \theta^{\alpha A}\big(\sigma^{\mu\nu}\big)_{\alpha\beta}
\theta^{\beta B}\epsilon_{AB}~~~~,
\label{theta}
\end{equation}
$F_{\mu\nu}^+$ is the self-dual part of the gauge field
strength and the dots in (\ref{chiralsuperfield}) stand for terms containing auxiliary
fields and derivatives.
The various components of the chiral superfield
are represented by the following open string vertex operators
\begin{subequations}
\label{gauge33}
\begin{align}
& V_A(z) = \frac{A_\mu(p)}{\sqrt 2}\,\psi^\mu(z)\,{\rm e}^{i p \cdot X(z)}\,
{\rm e}^{-\varphi(z)}~~,
\label{amu}
\\
& V_\Lambda(z)=
\Lambda^{\alpha A}(p)\,S_\alpha(z) S_{A}(z)\,{\rm e}^{i p\cdot X(z)}\,
{\rm e}^{-\frac{1}{2}\varphi(z)}~~,
\label{lambda}
\\
& V_{\phi}(z) = \frac{\phi(p)}{\sqrt 2}\,{\overline\Psi(z)}\,{\rm e}^{i p\cdot X(z)}\,
{\rm e}^{-\varphi(z)}~~,
\label{phi}
\end{align}
\end{subequations}
where $z$ is a world-sheet point, $p$ is the momentum along the D3 brane world-volume
($p^2=0$), and $\varphi$ is the boson of the superghost fermionization formulas (for
details, see Appendix \ref{secn:appA}).
For completeness, we also write the vertex operators for the
conjugate fields $\bar \Lambda_{\dot\alpha A}$ and $\bar \phi$,
namely
\begin{subequations}
\label{bargauge33}
\begin{align}
& V_{\bar\Lambda}(z) =
\bar\Lambda_{\dot\alpha A}(p)\,S^{\dot\alpha}(z)S^{A}(z)\,{\rm e}^{i p\cdot X(z)}\,
{\rm e}^{-\frac{1}{2}\varphi(z)}~~,
\label{barlambda}
\\
& V_{\bar\phi}(z) = \frac{\bar\phi(p)}{\sqrt 2}\,\Psi(z)\,{\rm e}^{i p\cdot X(z)}\,
{\rm e}^{-\varphi(z)}~~.
\label{phib}
\end{align}
\end{subequations}
In all these vertices, the polarizations have canonical
dimensions
\footnote{Unless explicitly mentioned we will always understand the appropriate
factors of $(2\pi\alpha')$, needed to have dimensionless
string vertices.} and are $[N]\times [N]$ matrices transforming
in the adjoint representation of $\mathrm{SU}(N)$ (here we neglect an overall factor of
$\mathrm{U}(1)$, associated to the center of mass of the $N$ D3 branes, which decouples and
does not play any r\^ole in our context).

By computing the field theory limit $\alpha'\to 0$ of all tree-level
scattering amplitudes among the vertex operators
(\ref{amu})-(\ref{phib}), one obtains
various couplings that lead to the
${\cal N}=2$ $\mathrm{SU}(N)$ super Yang-Mills action~\footnote{Compared to Ref.
\cite{Billo:2002hm}, here we have rescaled
all gauge fields with a factor of the
Yang-Mills coupling constant $g$ for later convenience.}
\begin{equation}
\begin{aligned}
&S_{\rm SYM}=
\int d^4x ~{\rm
Tr}\,\Big\{\frac{1}{2}\,F_{\mu\nu}^2 +2\, D_\mu \bar{\phi}\,D^\mu \phi
-2\,\bar\Lambda_{\dot\alpha A}\bar D\!\!\!\!/^{\,\dot\alpha \beta}
\Lambda_\beta^{\,A}\\
&~~~~~~~~~~+{\rm i}\sqrt 2 \,g\,\bar\Lambda_{\dot\alpha A}\epsilon^{AB}
\big[\,\phi, \bar\Lambda^{\dot\alpha}_{\,B}\big]
+{\rm i}\sqrt 2\,g\, \Lambda^{\alpha A}\epsilon_{AB}
\big[\,\bar \phi, \Lambda_{\alpha}^{\,B}\big]+g^2\big[\,\phi,
\bar\phi\,\big]^2~\Big\}~~.
\end{aligned}
\label{n1}
\end{equation}
In the following we will study the
non-perturbative structure of the ${\cal N}=2$ gauge effective action
when the chiral superfield acquires a v.e.v.
\begin{equation}
\langle \Phi_{uv} \rangle \equiv \langle \phi_{uv} \rangle =
a_{uv}=a_u\,\delta_{uv}
\label{vev}
\end{equation}
where $u,v=1,...,N$ and $\sum_u a_u=0$, so that the gauge group $\mathrm{SU}(N)$ is
broken to $\mathrm{U}(1)^{N-1}$. In particular we will investigate
non-perturbative instanton effects, also in presence of a non-trivial
supergravity background.

\vskip 0.6cm
\subsection{The instanton sector}
\label{subsec:inst}

In this stringy set-up instanton effects can be introduced by
adding $k$ D(--1) branes (or D-instantons) which give rise to
new types of excitations associated to open strings with at least one
end point on the D-instantons. Due to the Dirichlet boundary conditions
that prevent momentum in all directions, these new excitations
describe moduli rather than dynamical fields and are in one-to-one
correspondence with the ADHM
moduli of gauge instantons (for a more detailed discussion see, for instance,
Ref. \cite{Dorey:2002ik} and references
therein).

Let us consider first the open strings with both end-points on
the fractional D-instantons in the $\mathbb{Z}_2$ orbifold.
In this case, the NS sector contains six physical bosonic excitations that
can be conveniently organized in a vector $a'_\mu$ and a
complex scalar $\chi$, and also three auxiliary excitations $D_c$ ($c=1,2,3$).
The corresponding vertex operators are
\begin{subequations}
\label{vertNS}
\begin{align}
\label{verta}
&V_{a'}(z)= g_0\,a'_\mu \,\psi^{\mu}(z)\,\ee^{-\varphi(z)}~~,
\\
\label{vertchi}
&V_{\chi}(z)= \frac{\chi}{\sqrt 2}\,{\overline\Psi}(z)\,\ee^{-\varphi(z)}~~,
\\
\label{vertd}
&V_D(z)= \frac{D_c}{2}\,\bar\eta_{\mu\nu}^c\,
\psi^\nu(z)\psi^\mu(z)~~,
\end{align}
\end{subequations}
where $\bar\eta^c_{\mu\nu}$ are the three anti-self-dual 't Hooft symbols
, and $g_0$ is the D-instanton coupling constant
\begin{equation}
\label{g0gym}
g_0 = \frac{g}{4\pi^2\alpha'}~~.
\end{equation}
The R sector of the D(--1)/D(--1) strings contains instead
eight fermionic moduli, ${M}^{\alpha A}$ and
$\lambda_{\dot\alpha A}$, described by the following vertices
\begin{subequations}
\label{vertR}
\begin{align}
\label{vertM}
&V_{M}(z)= \frac{g_0}{\sqrt 2}\,{M}^{\alpha A} \,S_{\alpha}(z)
S_{A}(z)\,\ee^{-\frac{1}{2}\varphi(z)}~~,
\\
\label{vertla}
&V_{\lambda}(z)=
{{\lambda_{\dot\alpha A}}}\,S^{\dot\alpha}(z)S^{A}(z)
\,\ee^{-\frac{1}{2}\varphi(z)}~~.
\end{align}
\end{subequations}
All polarizations in the vertex operators (\ref{vertNS}) and (\ref{vertR})
are $[k]\times [k]$ matrices and transform in the adjoint representation
of $\mathrm{U}(k)$. In the following we will always understand the
$\mathrm{U}(k)$ indices for simplicity, unless they are needed to avoid ambiguities.
It is worth noticing that if the Yang-Mills coupling constant
$g$ is kept fixed when $\alpha'\to 0$ (as is appropriate to
retrieve the gauge theory on the D3 branes), then the dimensionful coupling $g_0$
in (\ref{g0gym}) blows up.
Thus, some of the vertex operators must be suitably rescaled with
factors of $g_0$ (like in (\ref{verta}) and (\ref{vertM}))
in order to yield
non-trivial interactions when $\alpha'\to 0$ \cite{Billo:2002hm}.
As a consequence some of the moduli acquire unconventional scaling dimensions
which, however, are
the right ones for their interpretation as parameters of an instanton solution
\cite{Dorey:2002ik,Billo:2002hm}. For instance, the
$a'_\mu$'s in (\ref{verta}) have dimensions of (length) and are related to the positions of the
(multi)-centers of the instanton, while
${M}^{\alpha A}$ in (\ref{vertM})
have dimensions of (length)$^{\frac{1}{2}}$ and are the fermionic
partners of the instanton centers. Furthermore, if we write the
$[k]\times[k]$ matrices ${a'}^\mu$ and ${M}^{\alpha A}$ as
\begin{equation}
\begin{aligned}
{a'}^\mu &= x_0^\mu\,\uno_{[k]\times[k]} + y^\mu_c\,T^c~~,
\\
{M}^{\alpha A}&=\theta^{\alpha A}\,\uno_{[k]\times[k]} + {\zeta}^{\alpha
A}_c\,T^c~~,
\end{aligned}
\label{xtheta}
\end{equation}
where $T^c$ are the generators of $\mathrm{SU}(k)$,
then the center of the instanton, $x_0^\mu$, and its fermionic partner, $\theta^{\alpha A}$,
can be identified respectively
with the bosonic and fermionic coordinates of the $\mathcal{N}=2$ superspace.

In the D3/D(--1) system there are also twisted string excitations
corresponding to open strings with mixed boundary conditions that
stretch between a D3 brane and a D-instanton, or vice versa.
The twisted NS sectors contains the bosonic moduli
$w_{\dot\alpha}$ and $\bar w_{\dot\alpha}$ which are
associated to the following vertex operators
\begin{equation}
\label{vertw}
V_w(z)= \frac{g_0}{\sqrt{2}}\,{w}_{\dot\alpha}\,
\Delta(z) S^{\dot\alpha}(z)\,\ee^{-\varphi(z)}~~~~{\rm and}~~~~
V_{\bar w}(z)= \frac{g_0}{\sqrt{2}}
\,{\bar w}_{\dot\alpha}\, \overline\Delta(z)
S^{\dot\alpha}(z)\, \ee^{-\varphi}(z)~~.
\end{equation}
Here $\Delta$ and $\overline\Delta$ are the
twist and anti-twist operators with conformal weight $1/4$
which change the boundary conditions of the longitudinal coordinates $X^\mu$ from
Neumann to Dirichlet and vice-versa by introducing a cut in the open-string
world-sheet \cite{Dixon:1985}.
The moduli $w_{\dot\alpha}$ and
$\bar w_{\dot\alpha}$ have dimension of (length) and are
related to the instanton size. The twisted R sector contains instead the fermionic moduli
$\mu^A$ and ${\bar\mu}^A$, with dimension of (length)$^{1/2}$,
described by the following vertex operators
\begin{equation}
\label{vertmu}
V_\mu(z)= \frac{g_0}{\sqrt{2}}\,{\mu}^A\,
\Delta(z) S_{A}(z)\, \ee^{-{\frac12}\varphi(z)}~~~~{\rm and}~~~~
V_{\bar\mu}(z)= \frac{g_0}{\sqrt{2}}\,{{\bar \mu}^A}\,
\overline\Delta(z) S_{A}(z)\, \ee^{-{\frac12}\varphi(z)}~~.
\end{equation}
In both (\ref{vertw}) and (\ref{vertmu}) the polarizations
transform in the bi-fundamental representations of the ${\mathrm U}(N)\times {\mathrm U}(k)$
group. Again, in most cases we will understand the corresponding
indices for simplicity.

Following the procedure explained in Ref. \cite{Billo:2002hm}, by computing all
tree-level interactions among the above vertex operators in the
limit $\alpha'\to 0$ with $g$ fixed (and hence with
$g_0\to\infty$) one can recover the ADHM action on the instanton
moduli space for the ${\cal N}=2$ theory, namely
\begin{equation}
S_{\rm moduli}=S_k^{\,\rm bos}+S_k^{\,\rm fer}+S_k^{\,\rm c}
\label{smoduli}
\end{equation}
with
\begin{subequations}
\begin{align}
\label{skbos}
&S_k^{\,\rm bos} ={\rm tr}_k\Big\{
-2\,[\chi^{\dagger},a'_\mu][\chi,{a'}^\mu] +
\chi^{\dagger}{\bar w}_{\dot\alpha} w^{\dot\alpha}\chi
+ \chi{\bar w}_{\dot\alpha} w^{\dot\alpha} \chi^{\dagger}\Big\}
\\
\label{skfer}
&S_k^{\,\rm fer} ={\rm tr}_k\Big\{{\rm i}\,
\frac{\sqrt 2}{2}\,{\bar \mu}^A \epsilon_{AB} \mu^B\chi^{\dagger}
-{\rm i}\,
\frac{\sqrt 2}{4}\,M^{\alpha A}\epsilon_{AB}[\chi^{\dagger},M_{\alpha}^{B}]
\Big\}
\\
\label{sconstr}
&S_k^{\,\rm c}={\rm tr}_k\Big\{\!-\ii D_c\big({W}^c +\ii
\bar\eta_{\mu\nu}^c \big[{a'}^\mu,{a'}^\nu\big]\big) \notag\\
&~~~~~~~~~~~~~~~~~-
\ii {\lambda}^{\dot\alpha}_{\,A}\big(\bar{\mu}^A{w}_{\dot\alpha}+
\bar{w}_{\dot\alpha}{\mu}^A  +
\big[a'_{\alpha\dot\alpha},{M'}^{\alpha A}\big]\big)\!
\Big\}
\end{align}
\end{subequations}
where
\begin{equation}
\label{defWc}
(W^c)_{ij} = w_{ui\dot\alpha}\,(\tau^c)^{\dot\alpha}_{~\dot\beta}
\, \bar w^{\dot\beta}_{~ju}
\end{equation}
with $u=1,...,N$ and $i,j=1,...,k$.
Since $D_c$
and $\lambda^{\dot\alpha}_{\,A}$ act as Lagrange multipliers, the term (\ref{sconstr})
yields the so-called bosonic and fermionic ADHM constraints
\begin{equation}
\label{constr}
\begin{aligned}
&{W}^c +\ii
\bar\eta_{\mu\nu}^c \big[{a'}^\mu,{a'}^\nu\big]=0~~,
\\
&\bar{\mu}^A{w}_{\dot\alpha}+
\bar{w}_{\dot\alpha}{\mu}^A  +
\big[a'_{\alpha\dot\alpha},{M'}^{\alpha A}\big]=0~~,
\end{aligned}
\end{equation}
while by varying $S_k^{\,\rm bos}$ and $S_k^{\,\rm fer}$
with respect to $\chi^\dagger$ and $\chi$ we
obtain the following equations
\begin{subequations}
\begin{align}
\frac{1}{2}\left\{\bar w_{\dot \alpha}w^{\dot \alpha},\chi\right\}
&+\left[a'_\mu,\left[a'^\mu,\chi\right]\right]
+\ii\,\frac{\sqrt 2}{4}\,\epsilon_{AB}
\left(\bar\mu^A\mu^B+M^{\alpha A}M_\alpha^{~B}
\right)
=0~~,
\label{eqchi}
\\
&\frac{1}{2}\left\{\bar w_{\dot \alpha}w^{\dot \alpha},\chi^\dagger\right\}
+\left[a'_\mu,\left[a'^\mu,\chi^\dagger\right]\right]=0~~.
\label{eqchidag}
\end{align}
\end{subequations}
It is interesting to observe the moduli action
(\ref{smoduli}) does not depend on the superspace coordinates
$x_0^\mu$ and $\theta^{\alpha A}$ defined in (\ref{xtheta}) and
that the quartic interaction terms of the bosonic part (\ref{skbos})
can be completely disentangled by means of dimensionless auxiliary
fields $Y_\mu$, $X_{\dot\alpha}$ and $\bar X_{\dot\alpha}$ (plus their
conjugate ones) which are associated to the following vertex
operators \cite{Billo:2002hm}
\begin{subequations}
\label{vertaux}
\begin{align}
&V_Y(z)={\sqrt 2} g_0 \,{Y_\mu}\,\overline\Psi(z)\,
\psi^\mu(z)~~~~~~~,~~~~\,
V_{Y^\dagger}(z)={\sqrt 2} g_0 \,{Y_\mu^\dagger}\,\Psi(z)\, \psi^\mu(z)~~,
\label{verty}
\\
&V_X(z)= g_0\, X_{\dot\alpha}\,\Delta(z) S^{\dot\alpha}(z)\,
\overline\Psi(z)~~~~,~~~~
V_{X^\dagger}(z)= g_0\, X_{\dot\alpha}^\dagger\,\Delta(z) S^{\dot\alpha}(z)\,\Psi(z)~~,
\label{vertx}
\\
&V_{\bar X}(z)= g_0\, \bar X_{\dot\alpha}\,\overline\Delta(z) S^{\dot\alpha}(z)\,
\overline\Psi(z)~~~~,~~~~
V_{\bar X^\dagger}(z)= g_0\, \bar X_{\dot\alpha}^\dagger(z)\,\overline\Delta(z)
S^{\dot\alpha}(z)\,\Psi(z)~~.
\label{vertxb}
\end{align}
\end{subequations}
The operators (\ref{verty}) describe excitations of the
D(--1)/D(--1) strings, while the vertices (\ref{vertx}) and
(\ref{vertxb}) account for states of the D3/D(--1) and D(--1)/D3
sectors respectively. Like any vertex associated to an auxiliary
field, also the vertices (\ref{vertaux}) can only be written in the 0 superghost
picture and are not BRST invariant.
Nevertheless, they can be safely used to compute scattering amplitudes
in the field theory limit. For example, let us consider the
3-point amplitude corresponding to the disk diagram of Fig. \ref{fig:dia1}\emph{a},
namely
\begin{equation}
\lvev V_{\bar X^\dagger}V_{w}V_{\chi}\rvev
\,\,\equiv\,\,C_{0}\!\int\frac{\prod_i dz_i}{dV_{\rm
CKG}}\,\,\times
\, \langle V_{\bar X^\dagger}(z_1)\, V_{w}(z_2)\,V_{\chi}(z_3)\rangle
\label{Xdaggerwchi1}
\end{equation}
where  $dV_{\rm CGKG}$ is the Conformal Killing Group volume element, and
$C_0$ is the normalization of any D(--1) disk amplitude
\cite{Billo:2002hm}
\begin{equation}
C_{0} = \frac{2}{(2\pi\alpha')^2}\frac{1}{g_0^2}\,=\,
\frac{8\pi^2}{g^2}
\label{C0}
\end{equation}
which is also the classical action of an instanton with charge
$k=1$.
Computing the correlation function among the vertex operators
using the OPE's reported in Appendix \ref{secn:appA} and reinstating in the polarizations
the appropriate factors of $(2\pi\alpha')$ (which cancel against those of $C_0$),
one easily finds that
\begin{equation}
\lvev V_{\bar X^\dagger} V_{w}  V_{\chi}\rvev
= -{\rm tr}_k \Big\{{\bar X}^{\dagger}_{\dot\alpha} \,w^{\dot\alpha}\chi
\Big\}~~.
\label{Xdaggerwchi2}
\end{equation}
Proceeding systematically in this way and computing all scattering
amplitudes involving the auxiliary vertices, we obtain a bosonic moduli
action with  cubic interaction terms only, namely
\begin{equation}
\begin{aligned}
{S\,'}_{k}^{\,\rm bos} &={\rm tr}_k \Big\{
2\, Y^{\dagger}_{\mu}Y^{\mu}+2\,Y^{\dagger}_{\mu}\,
\big[{a'}^\mu,\chi\big] +2\,Y_{\mu}\,\big[{a'}^\mu,\chi^{\dagger}\big]
\\
&+{\bar X}^{\dagger}_{\dot\alpha}X^{\dot\alpha}
+{\bar X}_{\dot\alpha}X^{\dagger\,\dot\alpha}
+{\bar X}^{\dagger}_{\dot\alpha} w^{\dot\alpha}\chi
+{\bar X}_{\dot\alpha} w^{\dot\alpha}\chi^{\dagger}
-\chi{\bar w}_{\dot\alpha} X^{\dagger\,\dot\alpha}
- \chi^{\dagger}{\bar w}_{\dot\alpha}X^{\dot\alpha}
\Big\}
\end{aligned}
\label{s'kbos}
\end{equation}
which is indeed equivalent to $S_k^{\,\rm bos}$ in (\ref{skbos})
after the auxiliary variables are integrated out.

\FIGURE{\centerline{
\psfrag{X}{$\bar X^\dagger$}
\psfrag{c}{$\chi$}
\psfrag{w}{$w$}
\psfrag{f}{$\phi$}
\psfrag{a}{\emph{(a)}}
\psfrag{b}{\emph{(b)}}
\includegraphics[width=0.8\textwidth]{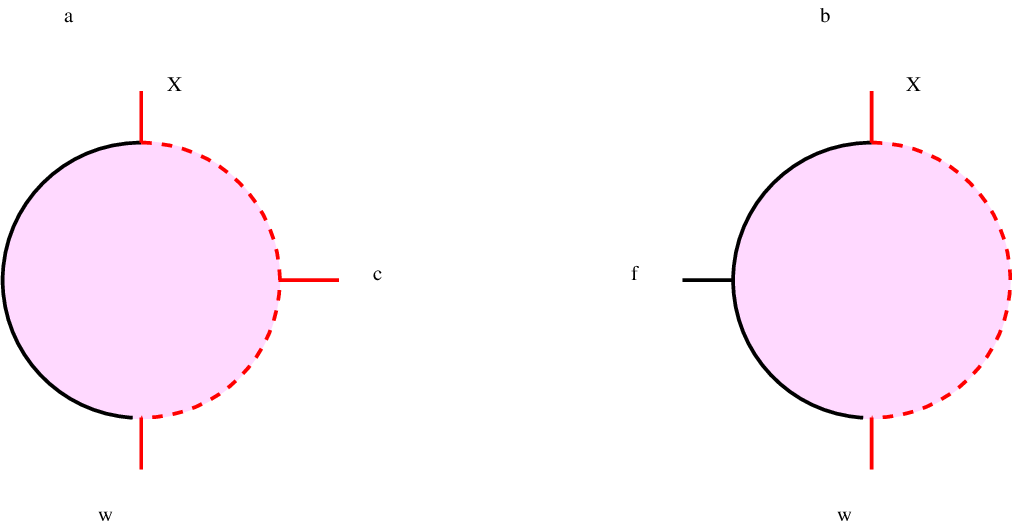}%
\caption{\emph{(a)} A mixed diagram involving moduli and auxiliary moduli.
\emph{(b)}
A mixed diagram involving also a
gauge theory vertex. The solid part of the boundary is attached to a D3,
the dotted part to a D-instanton.
\label{fig:dia1}}
}}

The vertices (\ref{vertaux}) are also useful to discuss the
supersymmetry transformation laws of the various moduli. In particular,
for the supersymmetries which are preserved both on the
D3 branes and on the D-instantons and which are generated by the
following four supercharges
\begin{equation}
Q^{\dot\alpha A} = \oint \frac{dw}{2\pi\ii}
\,j^{\dot\alpha A}(w)~~~~~~~{\rm with}~~~~~~
j^{\dot\alpha A}(w) = S^{\dot\alpha}(w) S^A(w)\,
\ee^{-\frac{1}{2}\varphi(w)}~~,
\label{susycharges}
\end{equation}
one can show that (see for example Ref. \cite{Billo:2002hm})
\begin{equation}
\comm{\xi_{\dot\alpha A}\, Q^{\dot\alpha A}}{V_Y(z)} =
\xi_{\dot\alpha A} \oint_z \frac{dw}{2\pi\ii}\, j^{\dot\alpha A}(w) \,
V_Y(z)
=V_{\delta M}(z)~~,
\label{auxYsusy}
\end{equation}
where
\begin{equation}
\delta M^{\beta B} = -2 {\sqrt 2}\,\xi_{\dot\alpha A}\,\epsilon^{AB}
({\bar\sigma}_\mu)^{\dot\alpha \beta}Y^\mu~~.
\label{auxsusy}
\end{equation}
After eliminating the auxiliary field
$Y$ via its equation of motion, one can rewrite
the above transformation rule as
\begin{equation}
\label{auxsusy2}
\delta M^{\beta B} = -2 {\sqrt 2}\,\xi_{\dot\alpha A}\,
\epsilon^{AB}
({\bar\sigma}_\mu)^{\dot\alpha \beta}\,[\chi,{a'}^\mu]~~.
\end{equation}
With similar calculations one can fully recover
also the supersymmetry transformations of the other instanton
moduli and find complete agreement with the standard results.

\vskip 0.6cm
\subsection{Introducing v.e.v.'s}
\label{subsecn:vevs}

The string formalism is well suited also to discuss the case in which
the chiral superfield $\Phi$ has a v.e.v. like in (\ref{vev}).
Indeed, to find how the constants $a_{uv}$ enter in the instanton action
one simply has to compute mixed disk diagrams with a
constant scalar field $\phi$ emitted from the portion of
the disk boundary that lies on the D3 branes. For example, one should
consider the diagram of Fig. \ref{fig:dia1}\emph{b}
which corresponds to the following amplitude
\begin{equation}
\lvev V_{\bar X^\dagger} V_{\phi} V_{w} \rvev
\,\,\equiv\,\,C_{0}\!\int\frac{\prod_i dz_i}{dV_{\rm
CKG}}\,\,\times
\, \langle V_{\bar X^\dagger}(z_1)\,V_{\phi}(z_2)\,V_{w}(z_3)\rangle
\label{Xdaggerwphi1}
\end{equation}
where the scalar vertex $V_\phi$ is taken at zero momentum to describe the emission of a
constant field $\phi$.
The amplitude (\ref{Xdaggerwphi1}) is similar to that of Eq.
(\ref{Xdaggerwchi1}), the only difference being
the presence of the D3/D3 vertex $V_\phi$ in place of the
D(--1)/D(--1) vertex $V_\chi$.
Since these vertices differ only in their
polarizations but not in their operator structure, the result can be simply inferred from
(\ref{Xdaggerwchi2}), namely
\begin{equation}
\lvev V_{\bar X^\dagger}  V_{\phi} V_{w}\rvev
= {\rm tr}_k \Big\{{\bar
X}^{\dagger}_{\dot\alpha}\,a \,w^{\dot\alpha}
\Big\}~~.
\label{Xdaggerwphi2}
\end{equation}
Proceeding systematically in this way, we can derive the modified
moduli actions
\begin{equation}
\label{skabos}
\begin{aligned}
{\tilde S}_{k}^{\,\rm bos} &={\rm tr}_k \Big\{
2\, Y^{\dagger}_{\mu}Y^{\mu}+2\,Y^{\dagger}_{\mu}\,
\big[{a'}^\mu,\chi\big] +2\,Y_{\mu}\,\big[{a'}^\mu,\chi^{\dagger}\big]
\\
&~~~~+{\bar X}^{\dagger}_{\dot\alpha}X^{\dot\alpha}
+{\bar X}_{\dot\alpha}X^{\dagger\,\dot\alpha}
+{\bar X}^{\dagger}_{\dot\alpha} \big( w^{\dot\alpha}\chi
- a\,w^{\dot\alpha}\big)
+{\bar X}_{\dot\alpha} \big(w^{\dot\alpha}\chi^{\dagger}
-{\bar a}\,w^{\dot\alpha}\big)
\\
&~~~~-\big(\chi{\bar w}_{\dot\alpha} -{\bar w}_{\dot\alpha}\,a\big)X^{\dagger\,\dot\alpha}
- \big(\chi^{\dagger}{\bar w}_{\dot\alpha}-{\bar w}_{\dot\alpha}\,\bar a\big)X^{\dot\alpha}
\Big\}~~,
\end{aligned}
\end{equation}
and
\begin{equation}
{\tilde S}_{k}^{\,\rm fer} ={\rm tr}_k\Big\{{\rm i}\,
\frac{\sqrt 2}{2}\,{\bar \mu}^A \epsilon_{AB} \big( \mu^B\chi^{\dagger}
-\bar a\,\mu^B\big)
-{\rm i}\,
\frac{\sqrt 2}{4}\,M^{\alpha A}\epsilon_{AB}[\chi^{\dagger},M_{\alpha}^{B}]
\Big\}~~.
\label{skafer}
\end{equation}
Notice that these actions can be obtained from those in Eqs. (\ref{s'kbos})
and (\ref{skfer}) with the formal shifts
\begin{equation}
\chi_{ij}\,\delta_{uv}\,\rightarrow\,
\chi_{ij}\,\delta_{uv}-\delta_{ij}\,a_{uv}
~~~~{\rm and}~~~~
\chi^\dagger_{ij}\,\delta_{uv}\,\rightarrow\,
\chi^\dagger_{ij}\,\delta_{uv}-\delta_{ij}\,{\bar a}_{uv}
\label{shift}
\end{equation}
where $i,j=1,...k$ and $u,v=1,...,N$. It is interesting to observe
that the matrices $a$ and $\bar a$ do not appear on equal
footing; in particular $a$ does not appear in the fermionic action
${\tilde S}_{k}^{\,\rm fer}$. This fact will have important consequences, like for example
that the instanton partition function depends only on $a$
and not on $\bar a$
(see also Section \ref{subsecn:holomo}).

\vskip 0.8cm
\section{Instantons in a graviphoton background}
\label{secn:gravi}

In this section we analyze the instanton
moduli space of $\mathcal{N}=2$ gauge theories in a non-trivial
supergravity background. In particular we turn on a (self-dual) field strength
for the graviphoton
of the $\mathcal{N}=2$ supergravity multiplet and see how it
modifies the instanton moduli action. This graviphoton background
breaks Lorentz invariance in space-time (leaving the metric flat) but
it allows to explicitly perform instanton calculations and
establish a direct correspondence with the
localization techniques that have been recently discussed in the
literature. Since our strategy is based on the use of
string and D brane methods, we begin by reviewing how the graviphoton field is
described in our stringy set-up.

\vskip 0.6cm
\subsection{Graviphoton in $\mathcal{N}=2$ theories}
\label{subsecn:graviphoton}

The graviton multiplet of $\mathcal{N}=2$ supergravity in four
dimensions contains the metric $g_{\mu\nu}$, two gravitini $\psi_\mu^{\,\alpha A}$
and one vector called graviphoton.
To describe the interactions of these fields with vector multiplets
it is convenient to introduce the
chiral Weyl superfield \cite{deWit:1984px}
\begin{equation}
W^+_{\mu\nu}(x,\theta)= \mathcal{F}_{\mu\nu}^+(x)+
\theta \chi_{\mu\nu}^+(x)+\frac{1}{2}
\,\theta\sigma^{\lambda\rho} \theta\,R^+_{\mu\nu\lambda\rho}(x)
+\cdots
\label{weyl}
\end{equation}
where the self-dual tensor
$\mathcal{F}_{\mu\nu}^+$ can be identified on-shell
with the graviphoton field strength,
$R^+_{\mu\nu\lambda\rho}$ is the
self-dual Riemann curvature tensor and
\begin{equation}
\theta \chi_{\mu\nu}\equiv\theta^{\alpha
A}\chi_{\mu\nu}^{~~\beta B}\,\epsilon_{\alpha\beta}\,\epsilon_{AB}
\label{gravitinofieldstrength}
\end{equation}
with $\chi_{\mu\nu}^{~~\alpha A}$
being the gravitino field strength, whose self-dual part appears in (\ref{weyl}).

In our context these supergravity fields are associated to massless
excitations of the type IIB closed string in $\mathbb{R}^{4}\times \mathbb{C}\times
\mathbb{C}^2/\mathbb{Z}_2$. Due to the presence of the fractional
branes, the closed string world-sheet has boundaries and
suitable identifications between left- and right-moving modes must be enforced.
Therefore a closed string vertex operator, which is normally the product
of two independent left and right components, {\it i.e.}
\begin{equation}
V_{\mathrm{L}}(z)\times V'_{\mathrm{R}}(\bar z)~~,
\label{vclosed}
\end{equation}
in the presence of D branes becomes of the form
\begin{equation}
V(z)\times V'(\bar z)
\label{vclosed1}
\end{equation}
where both the holomorphic and the anti-holomorphic
parts are written in terms of a single set of oscillators that
describe the modes of a propagating open string attached to the D branes.
Furthermore, due to these left/right identifications only eight of the
sixteen bulk supercharges that exist in the $\mathbb{Z}_2$
orbifold of Type IIB string theory
survive on the D brane world-volume.

Taking all this into account, we now write the vertex operators
associated to the fields of the $\mathcal{N}=2$ graviton
multiplet in the open string formalism. The graviphoton vertex operator belongs to
the R-R sector and in the $(-1/2,-1/2)$ superghost picture its
properly normalized expression is
\begin{equation}
V_{\mathcal{F}}(z,\bar z)=\frac{1}{4\pi}\,\mathcal{F}^{\alpha\beta AB}(p)\Big[
S_\alpha(z)S_A(z)\,\ee^{-\frac{1}{2}\varphi(z)}
\times {S}_\beta(\bar z){S}_B(\bar z)\,\ee^{-\frac{1}{2}{\varphi}(\bar
z)}\Big]\,\ee^{\ii p\cdot X(z,\bar z)}
\label{Vf}
\end{equation}
where the bi-spinor polarization is related to
$\mathcal{F}_{\mu\nu}^+$ in the following manner
\begin{equation}
\mathcal{F}^{\alpha\beta AB} = \frac{\sqrt 2}{4}\,
\mathcal{F}_{\mu\nu}^+\big(\sigma^{\mu\nu})^{\alpha\beta}\,\epsilon^{AB}~~,
\label{falphabeta}
\end{equation}
and
\begin{equation}
X^\mu(z,\bar z)=\frac 12\big[X^\mu(z)\pm X^\mu(\bar z)\big]
\label{xzbz}
\end{equation}
depending on whether $X^\mu$ is a longitudinal ($+$ sign) or
transverse ($-$ sign) direction~\footnote{To be very precise also
the overall sign of $V_{\mathcal{F}}$ (and of other closed string vertices)
depends on the type of boundary conditions; however, as we shall see in the following,
this sign is irrelevant in our calculations.}.

The vertex operator for the gravitini $\psi_\mu^{\alpha A}$
belongs instead to the fermionic R-NS/NS-R
sector and is given by
\begin{equation}
\begin{aligned}
V_{\psi}(z,\bar z)=\,&\frac{1}{4\pi}\,\psi_\mu^{\alpha A}(p)\,
\Big[S_\alpha(z)S_A(z)\,\ee^{-\frac{1}{2}\varphi(z)}\,\times\,
\psi^\mu(\bar z)\,\ee^{-\varphi(\bar z)}\\
&~~~~~~+\,
\psi^\mu(z)\,\ee^{-\varphi(z)}\,\times\,
{S}_\alpha(\bar z){S}_A(\bar z)\,\ee^{-\frac{1}{2}{\varphi}(\bar z)}\Big]\,
\ee^{\ii p\cdot X(z,\bar z)}~~.
\end{aligned}
\label{Vpsi}
\end{equation}
Notice that in the first term the holomorphic part is of R type with half-integer superghost
charge and the anti-holomorphic part is of NS type with integer superghost charge,
while the roles are reversed in the second term.
This ``symmetrized'' structure is a direct consequence of the
left/right identifications we have mentioned above.

Finally, the vertex operator for the graviton $h_{\mu\nu}$ belongs
to the NS-NS sector and in the $(-1,-1)$ picture it is
\begin{equation}
V_{h}(z,\bar z)= \frac{1}{4\pi}\,h_{\mu\nu}(p)\Big[
\psi^\mu(z)\,\ee^{-\varphi(z)}
\times \psi^\nu(\bar z)\,\ee^{-{\varphi}(\bar
z)}\Big]\,\ee^{\ii p\cdot X(z,\bar z)}~~~.
\label{Vh}
\end{equation}
The vertices (\ref{Vf}), (\ref{Vpsi}) and (\ref{Vh}) are,
as usual, dimensionless and their polarizations have canonical
dimensions. In particular, since the graviphoton field strength
$\mathcal{F}_{\mu\nu}$ has canonical dimensions of (length)$^{-1}$, a factor of
$(2\pi\alpha')^{1/2}$ should be understood in (\ref{Vf}).

Using the explicit expression of the above vertex operators and the OPE's given in
Appendix \ref{secn:appA}, it is possible to check various
supersymmetry transformation rules. For example, taking the
anti-chiral supercharges $Q^{\dot\alpha A}$ given in
(\ref{susycharges}), one can show that
\begin{equation}
\comm{\xi_{\dot\alpha A}\, Q^{\dot\alpha A}}{V_{\mathcal{F}}(z,\bar z)}
=V_{\delta\psi}(z,\bar z)
\label{VfVpsi}
\end{equation}
where
\begin{equation}
\delta \psi_\mu^{\,\beta B}= \ii\,\xi_{\dot\alpha A}\,(\bar
\sigma_\nu)^{\dot\alpha
\beta}\,\epsilon^{AB}\,\mathcal{F}_{\mu\nu}^+~~.
\label{supergrav}
\end{equation}
This is the correct graviphoton dependence of
the anti-chiral supersymmetry transformations of the gravitini.
Therefore, Eq. (\ref{VfVpsi})  is also a confirmation for the vertex operators (\ref{Vf})
and (\ref{Vpsi}). With similar calculations one can check other
pieces of the $\mathcal{N}=2$ supersymmetry transformation rules of
the various supergravity fields.

In the graviphoton vertex operator (\ref{Vf})
the holomorphic and anti-holomorphic components are both even under
the $\mathbb{Z}_2$ orbifold projection. For reasons that will be clear in the
following sections, it is convenient
to consider also a R-R closed string vertex
that is made up of two odd
components, namely
\begin{equation}
V_{\bar{\mathcal{F}}}(z,\bar z)=\frac{1}{4\pi}
\,\bar{\mathcal{F}}^{\alpha\beta \hat A\hat B}(p)\Big[
S_\alpha(z)S_{\hat A}(z)\,\ee^{-\frac{1}{2}\varphi(z)}
\times {S}_\beta(\bar z){S}_{\hat B}(\bar z)\,\ee^{-\frac{1}{2}{\varphi}(\bar
z)}\Big]\,\ee^{\ii p\cdot X(z,\bar z)}
\label{Vbarf}
\end{equation}
where $\hat A, \hat B=3,4$ in the notation of Appendix \ref{secn:appA}. This
vertex operator clearly survives the orbifold projection
since both $S_{\alpha}(z)S_{\hat A}(z)$ and $S_{\beta}(\bar z)S_{\hat B}(\bar z)$
are odd under $\mathbb{Z}_2$. In particular, we will consider the
case in which the bi-spinor polarization of $V_{\bar{\mathcal{F}}}$
is
\begin{equation}
\bar{\mathcal{F}}^{\alpha\beta \hat A\hat B} = \frac{\sqrt 2}{4}\,
\bar{\mathcal{F}}_{\mu\nu}^+\big(\sigma^{\mu\nu})^{\alpha\beta}\,\epsilon^{\hat A \hat B}
\label{barfalphabeta}
\end{equation}
where $\epsilon^{34}=-\epsilon^{43}=1$.
Notice that the antisymmetric tensor $\bar{\mathcal{F}}_{\mu\nu}$ cannot be interpreted
as the graviphoton field strength, since the vertex operator
(\ref{Vbarf}) is not related to the gravitino vertex (\ref{Vpsi})
as required by the rule (\ref{supergrav}) of $\mathcal{N}=2$ supersymmetry. In fact the
tensor $\bar{\mathcal{F}}_{\mu\nu}$ corresponds
to the field strength of some other vector in the $\mathcal{N}=2$
supergravity model, and as such it is independent of $\mathcal{F}_{\mu\nu}$.
Despite their different meaning,
the two vertices $V_{\mathcal{F}}$ and $V_{\bar{\mathcal{F}}}$
can be treated together in most of our calculations because of
their very similar operator structure.

\vskip 0.6cm
\subsection{ADHM measure in graviphoton background}
\label{subsecn:adhmgrav}

We now study how a graviphoton background modifies the instanton moduli action
of the $\mathcal{N}=2$ gauge theory. To do so, we assume that the chiral
graviphoton superfield (\ref{weyl}) has a v.e.v.
\begin{equation}
\langle W_{\mu\nu}^+ \rangle \equiv \langle \mathcal{F}_{\mu\nu}^+ \rangle = {f}_{\mu\nu}
\label{gravvev}
\end{equation}
with $f_{\mu\nu}$ a constant self-dual tensor. This
background can be described by a graviphoton vertex $V_{\mathcal{F}}$ at zero momentum
with constant polarization $\mathcal{F}_{\mu\nu}^+
=f_{\mu\nu}$, and the modified instanton action can be derived by computing
all disk amplitudes among the various moduli with
insertions of this closed string vertex.
These are mixed open/closed string amplitudes which are very similar to the ones
that have been previously studied in the context of non-anti-commutative
theories \cite{Billo:2004zq}.

\FIGURE{\centerline{
\psfrag{a}{\emph{(a)}}
\psfrag{b}{\emph{(b)}}
\psfrag{Y}{$Y^\dagger$}
\psfrag{f}{$\mathcal{F},\bar{\mathcal{F}}$}
\psfrag{fb}{$\bar{\mathcal{F}}$}
\psfrag{a1}{$a'$}
\psfrag{M}{$M$}
\includegraphics[width=0.8\textwidth]{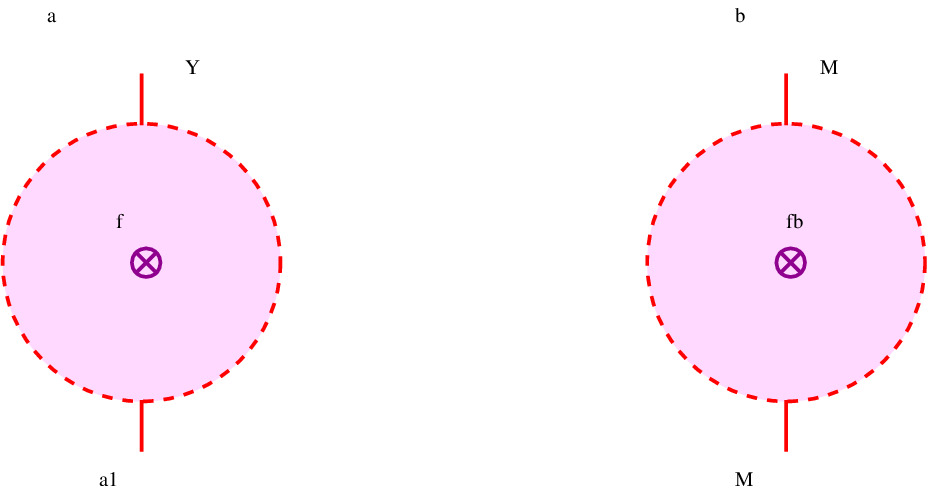}%
\caption{Disk diagrams encoding the coupling of the closed string RR
vertices $\mathcal{F}_{\mu\nu}$ (the graviphoton) and
$\bar{\mathcal{F}}_{\mu\nu}$ to bosonic \emph{(a)} and fermionic \emph{(b)} moduli.
The boundary of the disks is entirely on a D-instanton.}
\label{fig:disk_grav}
}}

Let us consider in detail the disk diagram
represented in Fig. \ref{fig:disk_grav}\emph{a}
which corresponds to the following amplitude:
\begin{equation}
\label{YaM}
\lvev V_{Y^\dagger} V_{a'}V_{\mathcal F} \rvev
\,\,\equiv\,\,C_{0}\!\int\frac{dz_1\, dz_2\, dw d\bar w}{dV_{\rm
CKG}}\,\,\times
\, \langle V_{Y^\dagger}(z_1)\, V_{a'}(z_2)\,V_{\mathcal F}(w,\bar w)\rangle
\end{equation}
where the open string punctures $z_i$ are integrated along the
real axis with $z_1\geq z_2$ while the closed string puncture
$w$ is integrated on the upper half complex plane.
More explicitly, we have
\begin{equation}
\begin{aligned}
\lvev &V_{Y^\dagger} V_{a'}V_{\mathcal F} \rvev
=\frac{1}{4\pi}\,{\rm tr}_k\Big\{Y^\dagger_\mu\,a'_\nu\,f_{\lambda\rho}
\Big\}\big(\sigma^{\lambda\rho}\big)^{\alpha\beta}\epsilon^{AB}\!\!
\int\frac{dz_1\,dz_2\, dw d\bar w}{dV_{\rm
CKG}}\,\times
\\
&~~\langle\,\ee^{-\varphi(z_2)}\ee^{-\frac{1}{2}\varphi(w)}
\ee^{-\frac{1}{2}\varphi(\bar w)}\rangle\,
\langle{\Psi}(z_1)S_{A}(w)S_{B}(\bar w)\rangle\,
\langle \psi^\mu(z_1)\psi^\nu(z_2) S_{\alpha}(w)
S_{\beta}(\bar w)\rangle~.
\end{aligned}
\end{equation}
Using the correlation functions given in Appendix \ref{secn:appA}
and exploiting the $\mathrm{Sl}(2,\mathbb{R})$ invariance to fix
$z_1\to\infty$ and $w\to\ii$, we are left with the elementary integral
\begin{equation}
\int_{-\infty}^{\infty}dz_2\,\frac1{1+z_2^2} = \pi~~,
\end{equation}
so that in the end we have
\begin{equation}
\lvev V_{Y^\dagger} V_{a'}V_{\mathcal F} \rvev
=-4\ii\,{\rm tr}_k\Big\{Y^\dagger_\mu\,a'_\nu\,f^{\mu\nu}
\Big\}~~.
\label{Yae}
\end{equation}
A systematic analysis reveals that this is the
only non-vanishing disk amplitude involving the graviphoton field strength $f_{\mu\nu}$
and
the ADHM instanton moduli. Indeed, all other diagrams with insertions of
$V_{\mathcal{F}}$ either
vanish at the string theory level, or vanish in the field theory limit.
However, there are a couple of non-vanishing amplitudes containing the vertex
$V_{\bar{\mathcal{F}}}$ of Eq. (\ref{Vbarf}) at zero momentum
({\it i.e.} with a constant polarization
${\bar{\mathcal{F}}}_{\mu\nu}^+=\bar f_{\mu\nu}$).
The first of these amplitudes, see Fig.  \ref{fig:disk_grav}\emph{a},
is the strict analogue of the one we have
presented above, namely
\begin{equation}
\lvev V_{Y} V_{a'}V_{\bar{\mathcal F}} \rvev
=-4\ii\,{\rm tr}_k\Big\{Y_\mu\,a'_\nu\,\bar f^{\mu\nu}
\Big\}~~.
\label{Yabare}
\end{equation}
The second is a fermionic amplitude involving the $M$ moduli,
see Fig. \ref{fig:disk_grav}\emph{b}, namely
\begin{equation}
\lvev V_{M} V_{M}V_{\bar{\mathcal F}} \rvev
=\frac{1}{4\sqrt 2}\,{\rm tr}_k\Big\{M^{\alpha A}\,M^{\beta B}\,\bar f_{\mu\nu}
\Big\}(\sigma^{\mu\nu})_{\alpha\beta}\epsilon_{AB}~~.
\label{mmbare}
\end{equation}
No other (irreducible) diagrams with $V_{\bar{\mathcal F}}$
insertions give a non-zero result. It is interesting to notice that the only
non-vanishing amplitudes with insertions of the closed string vertices
$V_{\mathcal F}$ and $V_{\bar{\mathcal F}}$ correspond to disks
whose boundary lies entirely on the D-instantons and that
there are no contributions to the instanton action due to graviphoton insertions on
mixed disks~\footnote{Things would be different in an
anti-self-dual graviphoton background.}.

By adding the contributions (\ref{Yae}), (\ref{Yabare}) and
(\ref{mmbare}) to the terms described in Section \ref{secn:N2} (in particular
Eqs. (\ref{skabos}) and (\ref{skafer})), we
can obtain the $\mathcal{N}=2$ ADHM moduli action in the presence of
a v.e.v. for the scalar field of the gauge multiplet, for the graviphoton field strength and
for the anti-symmetric tensor $\bar f_{\mu\nu}$. Explicitly, after integrating out the
auxiliary fields $Y$, $X$ and $\bar X$, the resulting moduli action is
\begin{equation}
\begin{aligned}
S_{\rm moduli}(a,\bar a;&f,\bar f) =\,-\,{\rm tr}_k\Big\{
2\,\big([\chi^{\dagger},a'_\mu]-2\,\ii\,\bar f_{\mu}^{~\nu} a'_\nu\big)
\big([\chi,{a'}^\mu]-2\,\ii\,f^{\mu\rho} a'_\rho\big) \\
&-
\big(\chi^{\dagger}{\bar w}_{\dot\alpha}-{\bar w}_{\dot\alpha}\,\bar a\big)
\big( w^{\dot\alpha}\chi- a\,w^{\dot\alpha}\big)
-\big(\chi{\bar w}_{\dot\alpha} -{\bar w}_{\dot\alpha}\,a\big)
\big(w^{\dot\alpha}\chi^{\dagger}
-{\bar a}\,w^{\dot\alpha}\big) \Big\}
\\
&+{\rm i}\,
\frac{\sqrt 2}{2}\,
{\rm tr}_k\Big\{{\bar \mu}^A \epsilon_{AB} \big(\mu^B\chi^{\dagger}
+\bar a\,\mu^B\big)
\\&
-\frac{1}{2}\,M^{\alpha A}\epsilon_{AB}\big([\chi^{\dagger},M_{\alpha}^{B}]
-\frac{\ii}{2}\,\bar
f_{\mu\nu}(\sigma^{\mu\nu})_{\alpha\beta}M^{\beta B}\big)
\Big\}+S_k^{\,\mathrm{c}}~~,
\end{aligned}
\label{saf}
\end{equation}
where $S_k^{\,\mathrm{c}}$ is the constraint part (\ref{sconstr})
which is not affected by the background
we have considered.

A few comments are in order at this point. First of all, if we
write the two self-dual tensors $f$ and $\bar f$ in terms of the three 't
Hooft's symbols
\begin{equation}
f_{\mu\nu}= f_c\,\eta^c_{\mu\nu}
~~~~{\rm and}~~~~
\bar f_{\mu\nu}= \bar f_c\,\eta^c_{\mu\nu}~~,
\label{fc}
\end{equation}
after some standard manipulations the action (\ref{saf}) becomes
\begin{equation}
\begin{aligned}
S_{\rm moduli}(a,\bar a;&f,\bar f) =\,-\,{\rm tr}_k\Big\{
\big([\chi^{\dagger},a'_{\alpha\dot\beta}]+2\,\bar f_c\,(\tau^c a')_{\alpha\dot\beta}\big)
\big([\chi,{a'}^{\dot\beta\alpha}]+2\,f_{c}\,(a'\tau^c)^{\dot\beta \alpha}\big) \\
&-
\big(\chi^{\dagger}{\bar w}_{\dot\alpha}-{\bar w}_{\dot\alpha}\,\bar a\big)
\big( w^{\dot\alpha}\chi- a\,w^{\dot\alpha}\big)
-\big(\chi{\bar w}_{\dot\alpha} -{\bar w}_{\dot\alpha}\,a\big)
\big(w^{\dot\alpha}\chi^{\dagger}
-{\bar a}\,w^{\dot\alpha}\big) \Big\}
\\
&+{\rm i}\,
\frac{\sqrt 2}{2}\,
{\rm tr}_k\Big\{{\bar \mu}^A \epsilon_{AB} \big( \mu^B\chi^{\dagger}
-\bar a\,\mu^B\big)
\\&
-\frac{1}{2}\,M^{\alpha A}\epsilon_{AB}\big([\chi^{\dagger},M_{\alpha}^{B}]
+2\,\bar f_c\, (\tau^c)_{\alpha\beta}M^{\beta B}\big)
\Big\}+S_k^{\,\mathrm{c}}~~,
\end{aligned}
\label{saf1}
\end{equation}
where, as usual,
$a'_{\alpha\dot\beta}=a'_\mu(\sigma^\mu)_{\alpha\dot\beta}$,
$a'^{\dot\beta\alpha}=a'_\mu(\bar\sigma^\mu)^{\dot\beta\alpha}$
and $\tau^c$ are the three Pauli matrices. When we use this
notation, it is clear that the effects on the instanton moduli of the gravitational
backgrounds $f$ and $\bar f$
can be formally introduced with the
following shifts
\begin{equation}
[\chi, (\bullet)_{\alpha}]\,\rightarrow\,
[\chi, (\bullet)_{\alpha}]
+2\, f_c \,(\tau^c\,\bullet)_{\alpha}
~~~{\rm and}~~~
[\chi^{\dagger}, (\bullet)_{\alpha}]\,\rightarrow\,
[\chi^{\dagger}, (\bullet)_{\alpha}]
+2\,\bar f_c \,(\tau^c\,\bullet)_{\alpha}
\label{shiftf}
\end{equation}
where the notation $(\bullet)_{\alpha}$ stands for any field in
the adjoint representation of $\mathrm{U}(k)$ that carries a
chiral Lorentz index $\alpha$. The shifts (\ref{shiftf}) are
in some sense the gravitational counterparts of the ones
in Eq. (\ref{shift}), which account for the presence of a non-trivial
v.e.v. for the gauge scalar fields, and appear as a rotation on the
chiral indices. The rules (\ref{shiftf}) very
much resemble the ones considered
in Refs. \cite{Nekrasov:2002qd,Flume:2002az,Losev:2003py,Flume:2004rp}
in the study of the localization properties of the instanton
moduli space.
Actually, we can be more precise in this respect.
In fact, by choosing the independent parameters $f_c$ and
$\bar f_c$ as
\begin{equation}
f_c=\frac{\varepsilon}{2}\,\delta_{3c}~~~~,~~~~
\bar f_c = \frac{\bar\varepsilon}{2}\,\delta_{3c}
\label{epsilon}
\end{equation}
with $\varepsilon=\bar \varepsilon$,
the action (\ref{saf1}) reduces exactly to the
one of Refs. \cite{Flume:2002az,Flume:2004rp}
with $\epsilon_1=-\epsilon_2=\varepsilon$. Thus, our derivation
gives a direct gravitational meaning to the deformation parameter
introduced in those references as a chiral rotation angle on some moduli;
in fact, in our context
this deformation naturally appears as due to a non-vanishing self-dual
graviphoton field strength of $\mathcal{N}=2$ supergravity.
In Section \ref{secn:omega}
we will provide more details on this point and also comment
on the relation of our results with the so-called
$\Omega$-background of Ref. \cite{Losev:2003py,Nekrasov:2005wp}.

\vskip 0.6cm
\subsection{Holomorphicity}
\label{subsecn:holomo}

Besides having a different meaning from the supergravity point of
view, the parameters $f_c$ and $\bar f_c$ are not
on equal footing in the instanton moduli action (\ref{saf1}).
In particular the graviphoton parameters $f_c$ do not appear
in the fermionic part of $S_{\rm moduli}$. As noticed at the end of
Section \ref{subsecn:vevs}, also the scalar v.e.v.'s $a_{uv}$ and $\bar a_{uv}$ have a
similar behaviour. This fact is not surprising since the effects of $a_{uv}$ and $f_c$
(or $\bar a_{uv}$ and $\bar f_c$) on the
instanton action can be generated by shifting $\chi$ (or $\chi^\dagger$)
according to the rules (\ref{shift}) and (\ref{shiftf}).
This structure has a very important consequence, namely the
instanton partition function
\begin{equation}
Z^{(k)}\equiv \int d\mathcal{M}_{(k)}\,\ee^{-S_{\mathrm{moduli}}(a,\bar a;f,\bar f)}
\label{partfunct}
\end{equation}
where $\mathcal{M}_{(k)}$ collectively denotes the instanton moduli,
depends only on $a_{uv}$ and $f_c$ and not
on $\bar a_{uv}$ and $\bar f_c$. Such holomorphicity is
a well-known property as far as the scalar v.e.v.'s are concerned \cite{Dorey:2002ik},
and here we extend it also to
the gravitational parameters $f_c$ and $\bar f_c$,
with a straightforward
generalization of the usual cohomology argument.

Let us give some details. In general, to prove holomorphicity it is rather convenient
to rearrange everything by means of the so-called topological twist.
This simply amounts to identify the  index
$A$ of the internal $\mathrm{SU}(2)_\mathrm{I}$ symmetry group with the anti-chiral index
$\dot \alpha$ of the $\mathrm{SU}(2)_\mathrm{R}$ factor of the Lorentz group.
After this identification, the new Lorentz group becomes
$\mathrm{SU}(2)_\mathrm{L}\times\mathrm{SU}(2)_\mathrm{d}$ where
$\mathrm{SU}(2)_\mathrm{d}$ is the diagonal subgroup of
$\mathrm{SU}(2)_\mathrm{R}\times\mathrm{SU}(2)_\mathrm{I}$.
The bosonic ADHM moduli $\{a'_\mu, \chi, \chi^\dagger,
w_{\dot\alpha}, \bar w_{\dot \alpha}, D_c\}$ are not affected by the twist,
while the fermionic ADHM moduli become
$\{\mu^{\dot\alpha},\bar\mu^{\dot\alpha},\eta,\lambda_c,M^\mu\}$
where
\begin{equation}
\lambda_c=\frac{\ii}{2}(\tau_c)^{\dot\alpha\dot\beta}\,\lambda_{\dot\alpha\dot\beta}
~~~,~~~\eta=\frac{1}{2}\,\epsilon^{\dot\alpha\dot\beta}\,\lambda_{\dot\alpha\dot\beta}
~~~,~~~
M^\mu=(\sigma^\mu)_{\alpha\dot\beta}\,M^{\alpha\dot\beta}~~.
\end{equation}
Similarly, the eight supersymmetry charges get reorganized as
$\{Q,Q_c,Q^\mu\}$ where
\begin{equation}
Q=\frac{1}{2}\,\epsilon_{\dot\alpha\dot\beta}\,Q^{\dot\alpha\dot\beta}~~~,~~~
Q_c=\frac{\ii}{2}(\tau_c)_{\dot\alpha\dot\beta}\,Q^{\dot\alpha\dot\beta}
~~~,~~~Q_\mu=(\sigma_\mu)^{\alpha\dot\beta}\,Q_{\alpha\dot\beta}
~~.
\end{equation}
The supercharge $Q$, which is the scalar component of the
four supercharges that are preserved both by the D3 and the D(--1)
branes (see Eq. (\ref{susycharges})), plays the role of a BRST
charge in the topologically twisted version of the $\mathcal{N}=2$
gauge theory. With some straightforward algebra, it is then possible to
show that the moduli action (\ref{saf1}) is $Q$-exact, namely
\begin{equation}
S_{\mathrm{moduli}}(a,\bar a;f,\bar f) = Q\,\Xi
\label{qxi}
\end{equation}
where
\begin{equation}
\begin{aligned}
\Xi~=~&{\rm tr}_k\Big\{\frac{1}{\sqrt 2}\Big(
\big(\chi^{\dagger}{\bar w}_{\dot\alpha}-{\bar
w}_{\dot\alpha}\,\bar a\big)\mu^{\dot\alpha}
+\bar\mu^{\dot\alpha}\big(w^{\dot\alpha}\chi^{\dagger}
-{\bar a}\,w^{\dot\alpha}\big)\Big)
\\
&-\sqrt 2 \,M_\mu\big([\chi^{\dagger},a'_\mu]-2\,\ii\,\bar f_{\mu}^{~\nu} a'_\nu\big)
-2\,\ii\,\lambda_c\big(\bar
w_{\dot\alpha}(\tau^c)^{\dot\alpha}_{~\dot\beta}w^{\dot\beta}
+\ii\bar\eta^c_{\mu\nu}[a'^\mu,a'^\nu]\big)\Big\}~~,
\end{aligned}
\label{xi}
\end{equation}
and the action of $Q$ on the various moduli is
\begin{equation}
\begin{aligned}
&Q\,a'^\mu=-\frac{\ii}{2}\,M^\mu~~,~Q\,\chi=0~~,~
Q\,\chi^\dagger=-\sqrt 2\,\ii\,\eta~~,~
\\
&Q\,D_c=-\,\ii\sqrt 2\,[\chi,\lambda_c]~~,~
Q\,M^\mu=\sqrt 2\big([\chi,{a'}^\mu]-2\,\ii\,f_{\mu\rho} a'^\rho\big)
~~,
\\
&Q\,\eta=\frac{1}{2}\,[\chi,\chi\dagger]
~~,~Q\,\lambda_c=\frac{1}{2}\,D_c~~,
\\
&Q\,w_{\dot\alpha}=-\frac{\ii}{2}\,\mu_{\dot\alpha}~~,~
Q\,\bar w_{\dot\alpha}=-\frac{\ii}{2}\,\bar\mu_{\dot\alpha}~~,
\\
&Q\,\mu^{\dot\alpha}=\sqrt 2\,\big(w^{\dot\alpha}\chi-a\,w^{\dot\alpha}\big)~~,~
Q\,\bar\mu^{\dot\alpha}=\sqrt 2\,\big(\chi\bar w^{\dot\alpha}
-w^{\dot\alpha}\,a)
~~.
\end{aligned}
\label{actionQ}
\end{equation}
It can be checked that $Q$ is indeed nilpotent (up to gauge transformations and
chiral rotations). Notice that $\bar a$ and $\bar f$ are present only
in the fermion $\Xi$ but not in the transformations (\ref{actionQ}).
On the contrary $a$ and $f$
appear explicitly through the action of $Q$. Therefore, making a
variation of the instanton partition function (\ref{partfunct})
with respect to $\bar a$ and $\bar f$
produces a $Q$-exact term and so $Z^{(k)}$ does not depend
on $\bar a$ and $\bar f$. For this reason, $\bar a$ and $\bar
f$ can be fixed to any convenient value.

\vskip 0.8cm
\section{Deformed ADHM construction and the $\Omega\,$-background}
\label{secn:omega}

As mentioned in the introduction, in supersymmetric models
the computation of the instanton partition function $Z^{(k)}$
for arbitrary $k$ is most easily performed if
the gauge theory is suitably deformed by turning on the so-called $\Omega$-background
\cite{Losev:2003py,Nekrasov:2005wp}.
On the instanton moduli space
this deformation corresponds to enlarge the symmetries of the ADHM construction.
In this section, following Ref. \cite{Flume:2004rp}, we briefly review this deformed
ADHM construction and show that it is directly related to the
instanton measure in a graviphoton background that we have
described in the previous section.

To this aim, let us introduce the
standard $[N+2k]\times [2k]$
ADHM matrix \footnote{Not to be confused with the twist field
$\Delta$ of Section \ref{secn:N2}.}
\begin{equation}
\label{salute1}
\Delta=\left(
\begin{array}{c} w \\
a^\prime-x\end{array}\right)~,
\end{equation}
where $w$ and $a'$ are a shorthand for $w_{ui\dot\alpha}$ and
$a'_{ij\alpha\dot\beta}$,
and
\begin{equation}
x \equiv\uno_{[k]\times [k]}\,\otimes\, \big(x_{\alpha\dot\beta}\big)=\uno_{[k]\times [k]}
\,\otimes\,\left(
\begin{array}{cc} z_1&-\bar z_2 \\
z_2&\bar z_1\end{array}\right)~,
\label{z}
\end{equation}
with $z_1$ and $z_2$ being the complex coordinates of the
(Euclidean) space-time. In the $\mathcal{N}=2$ theory
we introduce the fermionic partners of $\Delta$,
namely the
$[N+2k]\times [k]$ ADHM matrices
\begin{equation}
\label{salute10}
\mathcal{M}^A=\left(
\begin{array}{c}  \mu^A\\
M^A\end{array}\right)~,
\end{equation}
where $\mu^A$ and $M^A$ stand for $\mu_{ui}^{~~A}$ and
$M_{ij}^{~~\alpha A}$. In terms of these matrices, the bosonic and fermionic
ADHM constraints (\ref{constr}) read
\begin{equation}
\bar\Delta \Delta = \ell^{-1}\otimes \uno_{[2]\times [2]}~~~~\mathrm{and}~~~~
\bar\Delta\mathcal{M}^A+\bar{\mathcal{M}}^A\Delta=0~~,
\label{constr2}
\end{equation}
where $\ell$ is the $[k]\times [k]$ matrix
\begin{equation}
\ell =\big(\bar w_{\dot\alpha}w^{\dot\alpha}+(a'-x)^2\big)^{-1}~~.
\label{ell}
\end{equation}
Introducing a $[N+2k]\times [N]$ matrix $U$ such that
$\bar\Delta U = \bar U \Delta = 0$, the $\mathcal{N}=2$
$\mathrm{SU}(N)$ super-instanton
solution can be written as
\begin{subequations}
\begin{align}
&A_\mu=\frac{1}{g}\,\bar U\partial_\mu U~~,
\label{instamu} \\
&\Lambda^{\alpha A}=\frac{1}{g^{1/2}}\,
\bar{U}\left( {\mathcal M}^A \ell \,\bar{b}^\alpha
-b^\alpha \ell\, \bar{{\mathcal M}}^A\right)U~~,
\label{instlambda}\\
&\phi=\ii\,\frac{\sqrt 2}{4}\,\epsilon_{AB}\,
\bar{U}\,{\mathcal{M}}^A \ell\,\bar{\mathcal M}^B U
+\bar U\, \mathcal{J}\,U~,
\label{instphi}
\end{align}
\end{subequations}
where
$b^\alpha=\left(\begin{array}{c}
0\\
\delta^{\alpha}_{\beta}\,\delta_{ij}
\end{array}\right)
$
and $\mathcal{J}$ is the $[N+2k]\times [2k]$ matrix
\begin{equation}
\mathcal{J}=\left(\begin{array}{cc} 0
& 0
\\0
&\chi
\end{array}\right)~~.
\label{calJ}
\end{equation}
Here $\chi$ is a shorthand for $\chi \otimes\uno_{[2]\times[2]}$,
where $\chi$ is a hermitian
$[k]\times [k]$ matrix such that
\begin{equation}
\mathbf{L}\,\chi=-\ii\,\frac{\sqrt 2}{4}\,\epsilon_{AB}\,
\bar{\mathcal{M}}^A\mathcal{M}^B~,
\label{eqchi1}
\end{equation}
with the operator $\mathbf{L}$ defined by
\begin{equation}
{\bf L}\,
\bullet = \frac{1}{2}\{\bar w_{\dot \alpha}w^{\dot \alpha},\bullet\}
+[a'_\mu,[a'^\mu,\bullet]]~~.
\end{equation}
Notice that Eq. (\ref{eqchi1}) is the same equation (\ref{eqchi}) that follows
by varying the moduli action (\ref{smoduli}) with respect to $\chi^\dagger$.
In showing that Eq. (\ref{instphi}) is an instanton solution of
the scalar field equation, a crucial role is played by the
following zero-mode (see Appendix C of Ref. \cite{Dorey:2002ik})
\begin{equation}
D^\mu\big(\bar U\mathcal{J}\,U\big)=
2\,\mathrm{Im}\,\big(\bar U\mathcal{A}\,\bar\sigma^\mu\ell\bar
b\,U\big)~,
\label{cov1}
\end{equation}
where
\begin{equation}
\mathcal{A}=\left(\begin{array}{c}
-w \chi
\\
\left[\chi,a'\right]\end{array}\right)~~.
\label{aa0}
\end{equation}

This ADHM construction of the super-instanton solution can be
generalized to include a v.e.v. for the scalar field
$\phi$. In this case the classical profile of $\phi$
at the leading order in the Yang-Mills coupling
\footnote{\label{foot}
We assume the following expansions for the various gauge fields:
$A=g^{-1}A^{(0)}+g A^{(1)}+...~$;
$\Lambda=g^{-1/2}\Lambda^{(0)}+g^{3/2}\Lambda^{(1)}+...~$;
$\bar\Lambda=g^{1/2}\bar\Lambda^{(0)}+g^{5/2}\bar\Lambda^{(1)}+...~$;
$\phi=\phi^{(0)}+g^{2}\phi^{(1)}+...~$; $\bar\phi=\bar\phi^{(0)}+g^{2}\bar\phi^{(1)}+...~$.
At leading order for $g\to0$, one can just work with the first terms in these expansions.}
is still given by Eq. (\ref{instphi}), but with
\begin{equation}
\mathcal{J}\rightarrow \mathcal{J}(a)=\left(\begin{array}{cc} a
& 0
\\0
&\chi
\end{array}\right)~~,
\label{instphia}
\end{equation}
where $a$ is the $[N]\times[N]$ v.e.v. matrix and $\chi$ now satisfies
\begin{equation}
\mathbf{L}\,\chi=-\ii\,\frac{\sqrt 2}{4}\,\epsilon_{AB}\,
\bar{\mathcal{M}}^A\mathcal{M}^B +
\bar w_{\dot\alpha}\,a\,w^{\dot\alpha}~~.
\label{eqchi1a}
\end{equation}
Note that (\ref{eqchi1a}) is precisely the equation of motion that follows
varying w.r.t. to $\chi^\dagger$ the moduli
action presented in Section \ref{subsecn:vevs} (with $\bar a=0$).

The presence of a non-zero v.e.v. for $\phi$ can be interpreted as a deformation
of the ADHM construction. To appreciate this point, we can follow
Ref.\cite{Flume:2004rp} and show that the matrix $\mathcal{A}$
given in Eq. (\ref{aa0}) must be replaced by
\begin{equation}
\mathcal{A}(a)=\left(\begin{array}{c}
 a w-w \chi
\\
\left[\chi,a'\right]\end{array}\right)~~.
\label{aa}
\end{equation}
Notice that $\mathcal{A}(a)$ can be obtained from $\mathcal{A}$ by
means of
the same shift (\ref{shift}) that we have derived in Section
\ref{subsecn:vevs} from string amplitudes. The structure
(\ref{aa}) has also a further interpretation if we note that the ADHM
constraints are left invariant by the
transformations $T_\chi=\ee^{\ii\,\chi} \in \mathrm{U}(k)$ and
$T_{a}=\ee^{\ii\,a} \in \mathrm{SU}(N)$, which
reflect the redundancy in the ADHM description
and do not change the gauge connection (\ref{instamu}).
Under these transformations the ADHM data change as
\begin{equation}
\Delta ~\rightarrow~
\left(\begin{array}{c}
T_{ a}\, w\, T_{ \chi}^{-1}\\
T_{ \chi} \, (a^\prime-x)\, T_{ \chi}^{-1}
\end{array}\right) = \Delta +\,\ii\,\mathcal{A}(a) + \cdots
\label{ukun}
\end{equation}
Thus, the matrix (\ref{aa}) is related to the first order variation of
$\Delta$ under the symmetries of the ADHM
constraints.

This construction can be further generalized \cite{Flume:2004rp}
by including other symmetries of the ADHM constraints, in particular the
chiral rotations in the four dimensional Euclidean
space-time~\footnote{Actually, as discussed in Ref.
\cite{Flume:2004rp}, one could consider both chiral and
anti-chiral rotations of the space-time. However, for our purposes
it is enough to restrict our analysis to the chiral ones.} which, as shown in
Refs. \cite{Nekrasov:2002qd,Losev:2003py,Flume:2002az,Flume:2004rp},
allow to fully localize the integral on the instanton moduli space
on a discrete set of isolated fixed points.
Thus, in place of (\ref{ukun}) we consider
\begin{equation}
\Delta ~\rightarrow~
\left(\begin{array}{c}
T_{ a}\, w\, T_{ \chi}^{-1}\\
T_{ \chi} \, T_\varepsilon\,(a^\prime-x)\, T_{ \chi}^{-1}
\end{array}\right)
\label{ukunepsilon}
\end{equation}
where $T_{\varepsilon}=\ee^{\ii\,\varepsilon_c\tau^c}\in \mathrm{SU}(2)_{\mathrm{L}}$
generates chiral rotations in the complex $z_1$ and $z_2$ planes with angles
$\varepsilon_c^1=-\varepsilon_c^2=\varepsilon_c$. At first order we now have
$\Delta ~\rightarrow~
\Delta +\,\ii\,\mathcal{A}(a,\varepsilon)$,
where
\begin{equation}
\mathcal{A}(a,\varepsilon) =
\left(\begin{array}{c}
a w -w \chi
\\
\left[\chi,a'\right]+\varepsilon_c\,\tau^c a'
\end{array}\right)~~.
\label{aaepsilon}
\end{equation}
To find the instanton profile of the scalar field $\phi$ also
in the presence of the $\varepsilon$-rotations,
we take the Ansatz (\ref{cov1}) with $\mathcal{A}$ replaced by
$\mathcal{A}(a,\varepsilon)$. Since
\begin{equation}
\mathcal{A}(a,\varepsilon) =-\Delta \,\chi
+\left(\begin{array}{cc}a&0\\ 0& \chi+\varepsilon_c \,\tau^c\end{array}\right)\Delta
+\varepsilon_c\left(\begin{array}{c}0\\ \tau^c x\end{array}\right)~~,
\label{cala}
\end{equation}
we can show that \cite{Flume:2004rp}
\begin{equation}
2\,\mathrm{Im}\,\big(\bar U\mathcal{A}(a,\varepsilon)\,\bar\sigma^\mu\ell\bar
b\,U\big)=D^\mu\big(\bar U\mathcal{J}(a,\varepsilon)U\big)
-\ii\,g\,\Omega_{\nu\rho}\,x^\rho F^{\mu\nu}~,
\label{cov1aepsilon}
\end{equation}
where
\begin{equation}
\mathcal{J}(a,\varepsilon) =\left(\begin{array}{cc} a
& 0
\\0
&\chi
+\varepsilon_c
\tau^c\end{array}\right)~~,
\label{jaepsilon}
\end{equation}
\begin{equation}
\Omega_{\mu\nu}=\varepsilon_c\,\eta^c_{\mu\nu}~~,
\label{omega}
\end{equation}
and $F_{\mu\nu}$ is the instanton gauge field strength which
follows from (\ref{instamu}) and obeys $D^\mu F_{\mu\nu}=0$.
Furthermore, if $\chi$ satisfies the constraint
\begin{equation}
\mathbf{L}\,\chi=-\ii\,\frac{\sqrt 2}{4}\,\epsilon_{AB}\,
\bar{\mathcal{M}}^A\mathcal{M}^B +
\bar w_{\dot\alpha}\,a\,w^{\dot\alpha}-\ii\,\Omega_{\mu\nu}
\big[a'^\mu,a'^\nu\big]~~,
\label{eqchi1aepsilon}
\end{equation}
we can prove that Eq. (\ref{instphi}) with
$\mathcal{J}\rightarrow\mathcal{J}(a,\varepsilon)$
is a solution of the following field equation
\begin{equation}
D^2\phi = -\ii\,\sqrt 2\,g\,\epsilon_{AB}
\Lambda^{\alpha A}\Lambda_\alpha^{~B}- \ii\,g\, \Omega_{\mu\nu}
F^{\mu\nu}~~.
\label{eqphiomega}
\end{equation}

A few comments are in order. First of all, since the matrix
$\mathcal{A}(a,\varepsilon)$ is not homogeneous in $\Delta$ but contains also a
$x$-dependent piece proportional to $\varepsilon_c$, as is
clear from (\ref{cala}),
in computing $D^\mu\phi$ we produce also an extra piece proportional to
$\Omega$ that in turns modifies the structure of the scalar field
equation.
Secondly, the constraint (\ref{eqchi1aepsilon}) is precisely the
equation of motion for $\chi$ that follows by varying the moduli
action (\ref{saf}) in a constant graviphoton background with field
strength
\begin{equation}
f_{\mu\nu}=\frac 12\,\Omega_{\mu\nu}=\frac
12\,\varepsilon_c\,\eta^c_{\mu\nu}
\label{fomega}
\end{equation}
(and with $\bar a=\bar f_{\mu\nu}=0$).
Thus, our analysis shows that the parameters $\varepsilon_c$ of
the deformed ADHM construction of Ref. \cite{Flume:2004rp} have a direct interpretation in
terms of a constant graviphoton background, as we already
anticipated at the end of Section \ref{secn:gravi}.

At this point we can ask what is
the meaning of $\varepsilon_c$ at the level of the gauge theory action in four dimensions. In
Ref. \cite{Losev:2003py,Nekrasov:2005wp} it has
been argued that these deformation parameters are related to a non-trivial
metric in $\mathbb{R}^4\times\mathbb{C}$, called $\Omega$-background
and characterized by a self-dual antisymmetric tensor
$\Omega_{\mu\nu}$, which indeed leads to the
deformed field equation (\ref{eqphiomega}) at the leading order in
the Yang-Mills coupling constant. Here, however, we show that also on the
gauge theory the $\varepsilon$-deformation can be directly related to the
same graviphoton background (\ref{fomega}) that modifies the action on the
instanton moduli space.

To this aim, we first determine the deformed gauge theory action
by computing the couplings among the various gauge fields
and the graviphoton. This can be done by computing disk scattering amplitudes
among the vertex operators for the open string
massless excitations of the $N$ fractional D3 branes given in
(\ref{gauge33}) and (\ref{bargauge33}), and the closed string vertex operator
for the self-dual graviphoton field strength given in (\ref{Vf}).
For example, we have
\begin{equation}
\lvev V_{A} V_{\bar\phi}V_{\mathcal F} \rvev
=2\,\ii\,g\,\mathrm{Tr}\Big\{\partial_{[\mu}A_{\nu]}\bar\phi
f^{\mu\nu}\Big\}~~.
\label{abarphif}
\end{equation}
If we put $V_{\phi}$ instead of $V_{\bar\phi}$ we
get a vanishing result due to an unbalanced internal charge.
Actually, the amplitude (\ref{abarphif}) is the only non-zero 3-point function involving
the graviphoton field strength.

To find higher order contributions it is
convenient to follow the method described in detail
in Ref. \cite{Billo:2004zq}
in the context of non-anti-commutative theories and introduce the
auxiliary fields that disentangle the non-abelian quartic
interactions among the gauge vector bosons. It turns out that these auxiliary fields
have non-vanishing couplings also with the R-R graviphoton vertex
$V_{\mathcal F}$
and, when they are integrated out, two effects are obtained:
$\partial_{[\mu}A_{\nu]}$ in (\ref{abarphif}) is promoted to the full
non-abelian field strength $F_{\mu\nu}$, and a quartic term $\sim
g^2\,\big(\bar\phi f^{\mu\nu}\big)^2$ is produced.

Collecting all contributions, we find that
the action for the gauge fields of $N$ fractional D3 branes in
a self-dual graviphoton background $f_{\mu\nu}$ is given by
\begin{equation}
S_{\rm SYM}+\int d^4x ~{\rm
Tr}\,\Big\{-2\,\ii\,g\,F_{\mu\nu}\bar\phi
f^{\mu\nu}-g^2\big(\bar\phi f^{\mu\nu}\big)^2
\Big\}
\label{symf}
\end{equation}
where $S_\mathrm{SYM}$ is the super Yang-Mills action (\ref{n1}).
We remark that the deformation terms in (\ref{symf}) are in
perfect agreement with the general couplings between the Weyl and
gauge vector multiplets required by $\mathcal{N}=2$ supergravity
(see for instance the review \cite{Mohaupt:2000mj} and references
therein).
It is now easy to see that at the leading order in the coupling
constant $g$ (using the standard expansions mentioned in footnote \ref{foot}),
the field equation for $\phi$ that follows from (\ref{symf})
is precisely
Eq. (\ref{eqphiomega}) once the relation (\ref{fomega}) between
$f_{\mu\nu}$ and $\Omega_{\mu\nu}$ is taken into account.
Thus, the $f$-dependent terms in (\ref{symf}) correctly describe,
at the gauge theory level, the same deformation which on the instanton
moduli is realized as a chiral rotation.
The action (\ref{symf}) coincides with the
$\Omega$-background action of Refs. \cite{Losev:2003py,Nekrasov:2005wp}
at the linear order in $g$, but differs at higher orders. However,
since the instanton calculus is concerned with linearized actions, this difference
is unimportant.

Our analysis shows that both on the ADHM moduli space and on
the four-dimensional gauge theory the $\varepsilon$-rotations have
the same supergravity interpretation since they are related to the
components of the graviphoton field strength of the Weyl
multiplet.
Our method treats the $\varepsilon$-deformation in exactly the
same way on the ADHM moduli and on the gauge fields. This is quite
natural in the string realization of the instanton calculus by
means of systems of D3 and D(--1) branes, in which gauge fields
and ADHM moduli arise from different sectors of the same bound
state of D branes.

\vskip 0.8cm
\section{Instanton contributions to the effective action}
\label{secn:eff_action}

In this section we study the instanton partition function in a
graviphoton background. Using the holomorphicity properties
of Section \ref{subsecn:holomo} we can set $\bar a = \bar f_{\mu\nu}=0$
with no loss of generality, and concentrate only on the $a$ and $f_{\mu\nu}$ dependence.
These parameters are actually the v.e.v.'s of the lowest components of chiral superfields
but, implementing techniques and ideas of Ref. \cite{Green:2000ke}
(see also Ref. \cite{Billo:2002hm}), it is rather straightforward to
derive the full dependence on the
entire superfields $\Phi(x,\theta)$ (along the unbroken gauge directions of $\mathrm{U}(1)^{N-1}$)
and $W^+_{\mu\nu}(x,\theta)$.

\vskip 0.6cm
\subsection{The field-dependent moduli action}
\label{subsec:fdma}
Let us start by considering the gauge superfield $\Phi$. The
dependence of $S_{\mathrm{moduli}}$ on the lowest component $\phi$
can be derived by computing the same
mixed disk diagrams which produce the
$a$-dependent contributions, such as the one of Fig. \ref{fig:dia1}\emph{b},
but with a dynamical ({\it i.e.} momentum dependent) vertex $V_\phi$.
For example, the amplitude (\ref{Xdaggerwphi2})
becomes
\begin{equation}
\lvev V_{\bar X^\dagger} V_{\phi} V_{w} \rvev
= {\rm tr}_k \Big\{{\bar
X}^{\dagger}_{\dot\alpha}w^{\dot\alpha}\phi(p)
\,\ee^{\ii p\cdot x_0}\Big\}~~,
\label{Xdaggerwphi3}
\end{equation}
where the dependence on the instanton center $x_0$ (defined in
\eq{xtheta}) originates from the world-sheet correlator
$\vev{\bar\Delta(z_1)\, \ee^{\ii p\cdot X(z_2)} \Delta(z_3)} \propto \ee^{\ii p\cdot x_0}$.
A similar dependence on $x_0$ arises in all other mixed diagrams involving the vertex
operator $V_\phi$. From these string correlators, after taking the Fourier transform
with respect to $p$, we can extract a moduli action
which is given by \eq{saf} (at $\bar a=\bar f_{\mu\nu}=0$) with
\begin{equation}
\label{atophi}
a \rightarrow \phi(x_0)~~.
\end{equation}
With this replacement, $S_{\rm moduli}$, which originally did not depend
on the instanton center, acquires a non-trivial dependence on $x_0$
that, from now on, we will simply denote by $x$.

There are other non-vanishing disk diagrams that couple the various components of
the gauge supermultiplet to the instanton moduli.
These diagrams are related to the ones containing only $\phi$ by the
Ward identities of the supersymmetries that are broken by the D(--1)
branes and are generated by the chiral supercharges
\begin{equation}
\label{qbroken}
Q_{\alpha A} = \oint \frac{dw}{2\pi\ii}
\,j_{\alpha A}(w)~~~~~~~{\rm with}~~~~~~
j_{\alpha A} = S_{\alpha} S_A\,
\ee^{-\frac{1}{2}\varphi}~~.
\end{equation}
Note that the supercurrents $j_{\alpha A}$ (which carry trivial Chan-Paton factors) coincide
with the vertex for the moduli $\theta^{\alpha A}$ introduced in Eqs.\,(\ref{vertM})
and (\ref{xtheta}).
The transformations generated by
$\theta^{\alpha A} Q_{\alpha A}$ are precisely the ones
that connect the various components of the chiral
superfield $\Phi(x,\theta)$. For instance, we have
\begin{equation}
\label{susyL}
\comm{\theta^{\alpha A} Q_{\alpha A}}{V_\Lambda(z)} = V_{\delta \phi}(z)~~,
\end{equation}
where the vertices are given in Eqs.\,(\ref{lambda}) and (\ref{phi}), and
the supersymmetry variation
$\delta\phi = \theta^{\alpha A} \Lambda_{\alpha}^{~B}\epsilon_{AB}$
is encoded in the superfield structure of $\Phi(x,\theta)$, see \eq{chiralsuperfield}.
Thus, besides the amplitude (\ref{Xdaggerwphi3}) we also have a
correlator in which the vertex $V_\phi$ is replaced by
$V_{\delta\phi}$, {\it i.e.}
\begin{equation}
\label{sdisk}
\lvev V_{{\bar X}^\dagger} V_{\delta\phi} V_w \rvev =
\lvev V_{{\bar X}^\dagger} \comm{\theta^{\alpha A} Q_{\alpha A}}{V_\Lambda} V_w \rvev
~~.
\end{equation}
Deforming the integration contour for the supercharge
and taking into account the fact that $Q_{\alpha A}$
commutes with $V_{{\bar X}^\dagger}$ and
$V_w$, we can move it onto the D(--1) part of the boundary and get
\footnote{We refer to Ref. \cite{Green:2000ke} and in particular to Section 5.2 of the first
paper in Ref. \cite{Billo:2002hm} for more detailed explanations.}
\begin{equation}
\label{ward_id}
\begin{aligned}
\lvev V_{{\bar X}^\dagger} \comm{\theta^{\alpha A} Q_{\alpha A}}{V_\Lambda} V_w \rvev
= - \lvev V_{{\bar X}^\dagger} V_\Lambda V_w \!\int\! V_\theta \rvev~~.
\end{aligned}
\end{equation}
In this way
a 4-point amplitude containing one insertion of $V_\Lambda$ on the D3 boundary
and one (integrated) insertion of $V_\theta$ on the D(--1)
boundary can be related to a 3-point amplitude with $V_{\delta \phi}$. Therefore, the
corresponding result can be simply obtained from
\eq{Xdaggerwphi2} with the replacement
\begin{equation}
\label{phitotL}
\phi \to \delta\phi = \theta\Lambda~~.
\end{equation}
This analysis can be further iterated, revealing new couplings with the higher components
of $\Phi$ and more $\theta$-insertions. Altogether it turns out
that these additional interactions in the moduli action can be
summarized by extending the replacement (\ref{atophi}) to
\begin{equation}
\label{atoPhi}
a \rightarrow
\Phi(x,\theta)~~.
\end{equation}

A similar pattern can be followed also with the Weyl superfield (\ref{weyl}).
Introducing the momentum
dependence for the R-R vertex $V_{\mathcal{F}}$ in the amplitude
(\ref{YaM}), we obtain%
\footnote{Notice that the momentum dependence arises
from the one-point function of the
plane-wave term $\ee^{\ii p\cdot X(z,\bar z)}$ in the closed string
vertex $V_{\mathcal{F}}$ on a disk with D(--1) boundary conditions.}
\begin{equation}
\lvev V_{Y^\dagger} V_{a'}V_{\mathcal F} \rvev
=-4\ii\,{\rm tr}_k\Big\{Y^\dagger_\mu\,a'_\nu\,\mathcal{F}^{\mu\nu}(p)
\,\ee^{\ii p\cdot x}\Big\}~~.
\end{equation}
The couplings with the other components of $W^+_{\mu\nu}$ are related to this one by the
Ward identities for the supercharges $Q_{\alpha A}$ broken by the
D-instantons,
so that the dependence of $S_{\rm moduli}$ on
$W_{\mu\nu}^+$ can be obtained by simply replacing in \eq{saf}
\begin{equation}
\label{gravicouplings}
f_{\mu\nu} \rightarrow  W^+_{\mu\nu}(x,\theta)~~.
\end{equation}
Altogether, performing the replacements (\ref{atoPhi}) and (\ref{gravicouplings})
in $S_{\rm moduli}(a,0,f,0)$ given in \eq{saf}, we obtain
\begin{equation}
\label{sPW}
S_{\rm moduli}(a,0;f,0)\,\rightarrow\,
S_{\rm moduli}
\left(\Phi(x,\theta),0; W^+(x,\theta),0\right)
\equiv
S\big(\Phi; W^+; \widehat{\mathcal{M}}\big)~~,
\end{equation}
which describes the couplings of the abelian superfield $\Phi$ in the unbroken gauge directions
and of the Weyl superfield $W^+_{\mu\nu}$ to the centered instanton moduli
$\widehat{\mathcal{M}}$, {\it i.e.} all moduli except $x$ and $\theta$.

\vskip 0.6cm
\subsection{The low-energy effective action and the prepotential}
\label{subsec:leea}

Let us neglect for the moment the Weyl multiplet and concentrate on the
usual $\mathcal{N}=2$ SYM theory for which
the low-energy effective action is
a functional of the chiral superfield $\Phi$ in the unbroken gauge
directions
(and of its conjugate $\bar\Phi$). For simplicity, but without loss of generality,
we focus on the $\mathrm{SU}(2)$ gauge theory; in this case, from
\eq{vev} we see that the unbroken $\mathrm{U}(1)$ chiral
superfield is $\Phi=\Phi_3\,\tau^3$. In the following $\Phi_3$ will be simply denoted by
$\Phi$ without any ambiguity.

Up to two-derivative terms,
$\mathcal{N}=2$ supersymmetry constrains
the effective action for $\Phi$ to be of the form
\begin{equation}
\label{Seff}
S_{\mathrm{eff}}[\Phi] = \int d^4x \,d^4\theta\, \mathcal{F}(\Phi) + \mathrm{c.c}~~,
\end{equation}
where $\mathcal{F}$ is the prepotential. In the
semi-classical limit
$\mathcal{F}$ displays a 1-loop perturbative contribution plus instanton
corrections \cite{Seiberg:1988ur}, namely
\begin{equation}
\label{Fexp}
\mathcal{F}(\Phi) = \frac{\ii}{2\pi} \Phi^2 \log \frac{\Phi^2}{\Lambda^2} +
\sum_{k=1}^\infty \mathcal{F}^{(k)}(\Phi)~~,
\end{equation}
where $\Lambda$ is the dynamically generated scale and $k$ is the instanton number.

Focusing on a given sector with positive $k$, the instanton induced effective action for
$\Phi$ is given by \footnote{Since we are dealing with
$\mathcal{N}=2$ theories, we do not distinguish between effective
actions a la Wilson and effective actions a la Coleman-Weinberg.}
\begin{equation}
\label{seff1}
S_{\mathrm{eff}}^{(k)}[\Phi] = \int d^4x \, d^4\theta\,
d\widehat{\mathcal{M}}_{(k)}\, \,\ee^{-\frac{8\pi k}{g^2} \,-\, S
\left(\Phi;\widehat{\mathcal{M}}_{(k)}\right)}~,
\end{equation}
where in the exponent we have added also the classical part
of the instanton action $S_k^{\rm cl} = 8\pi k/g^2$. Upon
comparison with \eq{Seff}, we obtain
\begin{equation}
\label{cipf}
\mathcal{F}^{(k)}(\Phi) = \int d\widehat{\mathcal{M}}_{(k)}\,\,
\ee^{-\frac{8\pi k}{g^2} \,-\,
S\left(\Phi;\widehat{\mathcal{M}}_{(k)}\right)}~~.
\end{equation}
Thus, the $k$-instanton contribution to
the prepotential is given by the centered $k$-instanton
partition function \cite{Fucito:1996ua,Hollowood:2002ds}. Since the superfield $\Phi(x,\theta)$
is a constant with respect to the integration variables
$\widehat{\mathcal{M}}_{(k)}$, one can compute $\mathcal{F}^{(k)}$ by
fixing $\Phi(x,\theta)$ to its v.e.v., use the existing
results of the literature (see for example Ref. \cite{Dorey:2002ik})
and finally replace the v.e.v. with the
complete superfield $\Phi(x,\theta)$.
In this way one finds
\begin{equation}
\label{Fkform}
\mathcal{F}^{(k)}(\Phi) = c_k \, \Phi^2\, \left(\frac{\Lambda}{\Phi}\right)^{4k}~~.
\end{equation}
where the factor $\Lambda^{4k}$ originates from the
classical action term $\exp(-8\pi k/g^2)$, upon taking into
account the $\beta$-function of the
$\mathcal{N}=2$ $\mathrm{SU}(2)$ theory. The numerical coefficients $c_k$ have been
explicitly computed for $k=1$ and $k=2$ by evaluating the integral
over the instanton moduli space \cite{Dorey:1996hu,Fucito:1996ua} and checked against the
predictions of the Seiberg-Witten theory \cite{Seiberg:1994rs}, finding perfect
agreement. More recently, using the localization formulas of the
instanton integrals, the coefficients $c_k$ have been computed also for
arbitrary $k$ \cite{Nekrasov:2002qd,Flume:2002az,Klemm:2002pa,Flume:2004rp}.

Let us now consider the gravitational corrections to the $\mathcal{N}=2$ effective theory
by introducing also the dependence on the
Weyl superfield $W^+_{\mu\nu}$ induced by the D-instantons.
In perfect analogy with Eqs. (\ref{seff1})
and (\ref{cipf}), we have to construct connected diagrams that describe
also the couplings of the instanton moduli to $W_{\mu\nu}^+$,
and thus write
\begin{equation}
\label{seff2}
S_{\mathrm{eff}}^{(k)}[\Phi; W^+] = \int d^4x \, d^4\theta\,\mathcal{F}^{(k)}(\Phi; W^+)
\end{equation}
where the prepotential is
\begin{equation}
\label{cipfw}
\mathcal{F}^{(k)}(\Phi; W^+) = \int d\widehat{\mathcal{M}}_{(k)}\,
\ee^{-\frac{8\pi k}{g^2} \,-\,
S\left(\Phi; W^+; \widehat{\mathcal{M}}_{(k)}\right)}~~.
\end{equation}
Since both $\Phi(x,\theta)$ and $W_{\mu\nu}^+(x,\theta)$ are constant with respect to the
integration variables, we can simply compute $\mathcal{F}^{(k)}(a;f)$ and then
replace the v.e.v.'s with the corresponding superfields in the result.
By examining the explicit form of the moduli action $S_{\rm moduli}(a,0;f,0)$
given in \eq{saf}, we see that
it is invariant under the simultaneous sign reversal of $a$ and
$f$, if at the same time also the signs of $\chi$, $w$ and of the $\mathrm{SU}(k)$ part of
$a'$ (named $y_c$ in \eq{xtheta}) are reversed.
This is a change of integration variables in \eq{cipfw} with unit
Jacobian, so that we can conclude that
$\mathcal{F}^{(k)}(a;f)$ is invariant under the exchange
\begin{equation}
\label{af_sign}
a, f_{\mu\nu} \to -a, -f_{\mu\nu}~~.
\end{equation}
The prepotential $\mathcal{F}^{(k)}(a;f)$
has a regular expansion for $f\to 0$, where it reduces to the super Yang-Mills expression
$\mathcal{F}^{(k)}(a)$ of \eq{Fkform}. Moreover, it cannot contain
odd powers of $(a f_{\mu\nu})$, that would be compatible with the symmetry
(\ref{af_sign}) but would have necessarily some uncontracted indices and
therefore are unacceptable because of their tensorial nature. As a consequence, the expansion
of $\mathcal{F}^{(k)}(a;f)$ must contain even powers of both $a$ and $f_{\mu\nu}$,
the latter suitably contracted.
Replacing the v.e.v.'s with the corresponding
superfields, and remembering that the prepotential has dimensions of
(length)$^{-2}$, from the previous arguments we can deduce that
\begin{equation}
\label{Fkpw}
\mathcal{F}^{(k)}(\Phi; W^+) =  \sum_{h=0}^\infty c_{k,h} \,\Phi^2
\left(\frac{\Lambda}{\Phi}\right)^{4k}\!\left(\frac{W^+}{\Phi}\right)^{2h}~~,
\end{equation}
where $c_{k,0}=c_k$ so as to reproduce \eq{Fkform} at $W^+=0$. The
problem of finding the non-perturbative gravitational
contributions to the $\mathcal{N}=2$ superpotential is then
reduced to that of finding the numerical coefficients $c_{k,h}$.

The series (\ref{Fkpw}) is obtained from a perturbative expansion of the
linear couplings to the graviphoton multiplet
which appear in $\ee^{-S(\Phi;W^+;\mathcal{M}_{(k)})}$.
As discussed in
detail in Section \ref{subsecn:adhmgrav}, these couplings originate from disk diagrams
whose boundary lies on the D(--1) branes, each of which emits a graviphoton.
Thus, the term of order $(W^+)^{2h}$ in \eq{Fkpw}
comes from $2h$ disks, which correspond to a single (degenerate) Riemann surface with $2h$ boundaries
and Euler characteristic
\begin{equation}
\label{euler}
\chi_{\rm Euler} = 2 h - 2~~.
\end{equation}
This world-sheet is seemingly disconnected, but
in the construction of the effective action it plays
the r\^ole of a {\em connected} diagram
because of the integration over the instanton moduli (see the related discussion in Section 6
of Ref.%
\cite{Billo:2002hm}).

Let us consider now the entire set of non-perturbative contributions to the
prepotential
\begin{equation}
\label{Fnp}
\mathcal{F}_{\rm n.p.}(\Phi; W^+) = \sum_{k=1}^\infty \mathcal{F}^{(k)}(\Phi; W^+)
= \sum_{h=0}^\infty C_h(\Lambda,\Phi) (W^+)^{2h}~~,
\end{equation}
where
\begin{equation}
\label{Cg}
C_h(\Lambda,\Phi) = \sum_{k=1}^\infty c_{k,h} \,\frac{\Lambda^{4k}}{\Phi^{4k+2h-2}}~~.
\end{equation}
The effective action
corresponding to this prepotential contains many different terms, connected to each other by
supersymmetry. For instance, if we saturate the $\theta$-integral with four $\theta$'s
all coming from the $\Phi$ superfield, we obtain, among others,
four-fermion contributions proportional to
\begin{equation}
\Lambda^{4k} \int d^4x \,\,\phi^{-4k-2h-2} \big(\Lambda^{\alpha
A}\Lambda_{\alpha}^{~B}\epsilon_{AB}\big)^2\,\big(\mathcal{F}^+\big)^{2h}~~,
\label{fourfer}
\end{equation}
which for $h\not=0$ represent the gravitational corrections to the
four-gaugino interaction
induced by an instanton of charge $k$.

If instead the $\theta$-integral is saturated with four $\theta$'s
all coming from the $W^+$ superfields, we obtain, among others,
a contribution proportional to
\begin{equation}
\label{R2W}
\int d^4x
\,\,C_h(\Lambda,\phi)\,(R^+)^2  (\mathcal{F}^{+})^{2h-2}~~.
\end{equation}
When the scalar field $\phi$ is frozen to its expectation value,
this describes a purely gravitational F-term of the $\mathcal{N}=2$ effective action.
As we remarked above, in our approach based on the instanton
calculus, the contribution (\ref{R2W}) arises from mixed open/closed
string amplitudes on $2h$ disks.
Originally, this structure was discovered by computing pure closed
string amplitudes on Riemann surfaces of genus $h$
and through them a precise connection with the topological string was
established. We will comment more on this point in the following
subsection.

\vskip 0.6cm
\subsection{Relation with topological amplitudes}
\label{subsec:rel_top}

The gravitational F-terms (\ref{R2W}) can be computed also
in the context of Type II strings compactified on a Calabi-Yau (CY)
manifold, which is another well-known string setting from which one can
obtain a $\mathcal{N}=2$ low-energy effective theory.
In this case, the coefficients $C_h$
are functions of the moduli of the CY manifold \cite{Bershadsky:1993cx,Antoniadis:1993ze}
which can be computed by topological $h$-loop string amplitudes arising
from closed world-sheets of genus $h$, whose
Euler characteristic is given again by \eq{euler}.
When the two settings correspond to the same effective theory,
the topological string computation of $C_h$ should be in
agreement with the gauge theory instanton calculations presented
here, and indeed this is the case.

As shown by Seiberg and Witten (SW) \cite{Seiberg:1994rs},
the $\mathcal{N}=2$ low-energy effective action for a super Yang-Mills
theory with a gauge group $G$ can be described in terms of an
auxiliary Riemann surface. The so-called geometrical engineering
constructions \cite{Kachru:1995fv,Katz:1996fh}
embed the $\mathcal{N}=2$ theory in a consistent Type IIB string context, thus
accounting for the ``physical'' emergence of the SW Riemann surface.
For this one has to consider Type IIB strings on a ``local'' CY manifold $\mathcal{M}_G^{(B)}$
whose geometrical moduli are related to the quantities characterizing the gauge theory,
namely the dynamically generated scale $\Lambda$ and the gauge invariant composites
$\Tr\,\Phi^k$ of the scalars
\footnote{The manifold $\mathcal{M}_G^{(B)}$ is usually determined via
``local mirror symmetry'' \cite{Katz:1996fh} from a type IIA CY
manifold $\mathcal{M}_G^{(A)}$, whose
singularity structure reproduces the $\mathcal{N}=2$ effective
theory for $G$, in the low-energy limit
and upon decoupling gravity.
The form of the IIB local CY space $\mathcal{M}^{(B)}_G$ can also be
inferred directly, independently of the mirror construction,
as explained for example in Ref. \cite{denef_phd}.}.
The dependence of the IIB prepotential on the moduli of
$\mathcal{M}^{(B)}_G$ has a geometric expression in terms of periods
of suitable forms. Mapping the geometrical moduli to gauge theory quantities,
the prepotential matches
the field-theoretic SW expression \cite{Kachru:1995fv,Katz:1996fh}.
Also the  higher genus topological \cite{Bershadsky:1993cx,Antoniadis:1993ze} amplitudes $C_h$
on $\mathcal{M}^{(B)}_G$ can be computed as functions of the moduli
\cite{Chiang:1999tz,Klemm:2002pa}, and hence
of the gauge theory parameters.
In this way one can get, for instance, explicit expressions
for the couplings $C_h(\Lambda,\Phi)$  of Eq. (\ref{Cg})
in the $\mathrm{SU}(2)$ case.

As we argued above, the non-perturbative
superpotential $\mathcal{F}_{\mathrm{n.p.}}(\Phi;W^+)$
can be computed also on the gauge theory side, where it is given by the multi-instanton
centered partition functions (\ref{cipfw}), by freezing $\Phi$
and $W^+_{\mu\nu}$ to their v.e.v.'s $a$ and $f_{\mu\nu}$.
We have shown in Section \ref{subsecn:adhmgrav} (see in particular
Eqs. (\ref{fc}) and (\ref{epsilon})),
that the choice $f_{\mu\nu}= \frac{1}{2}\,\varepsilon\,\eta^3_{\mu\nu}$,
corresponds to the deformation of the instanton moduli space with
parameter $\varepsilon$ that has been introduced in Refs.
\cite{Nekrasov:2002qd,Flume:2002az,Flume:2004rp}
as a tool for the explicit evaluation
of multi-instanton partition functions.
Therefore, the expansion in powers of the
deformation parameter $\varepsilon$ and of the v.e.v. $a$ of the
centered multi-instanton partition functions
determines the coefficient $c_{k,h}$ of Eq. (\ref{Cg}).

That these coefficients must agree with those derived from the
topological amplitudes on $\mathcal{M}^{(B)}_G$ is a conjecture
put forward in  Ref. \cite{Nekrasov:2002qd} and checked
in Ref. \cite{Klemm:2002pa} for the $\mathrm{SU}(2)$ case.
We think that in the present paper we have made this conjecture extremely natural and
self-evident by recognizing that the deformation parameter
is nothing else that the graviphoton itself. Furthermore, our
analysis puts the evidence found in Ref.
\cite{Klemm:2002pa} in a broader perspective.

\vskip 0.6cm
\subsection{Consequences of the $\varepsilon$-deformation for the prepotential}
\label{subsecn:comp_loc}

There is however an important subtlety to be considered in checking the agreement
between the instanton calculations and the topological string results. To reproduce
the deformations (\ref{epsilon}), besides the graviphoton background
we have to turn on also a vacuum expectation value
$\bar f_{\mu\nu}= \frac{1}{2}\,\bar\varepsilon\, \eta^3_{\mu\nu}$ for a different R-R
field strength (setting moreover $\bar\varepsilon=\varepsilon$
as discussed in Section \ref{subsecn:adhmgrav}).
The holomorphicity properties of the
instanton moduli action discussed in Section \ref{subsecn:holomo}
ensure that the instanton partition
function does not smoothly depend of $\bar \varepsilon$; however, the
case $\bar\varepsilon=0$ is a limiting one and some care is needed.

To determine the coefficients $c_{k,h}$ of the prepotential expansion, it is enough to
consider constant background values for the scalar and Weyl
multiplets. In this case, the instanton contributions to the prepotential
$\mathcal{F}^{(k)}(a;f)$ of \eq{cipfw}
are well defined, but of course the corresponding contributions to the effective action
$S^{(k)}[a;f]$ diverge because of the (super)volume integral $\int d^4x \,d^4\theta$.
However, in presence of the complete deformations
(\ref{epsilon}) ({\it i.e.} when also $\bar \varepsilon$ is present),
the superspace integral is regularized by a gaussian term, and
thus it becomes possible to work at the level of the effective action,
that is at the level of the {\em full} instanton partition function
(as opposed to the centered one).
To be concrete, let us consider the simplest case $k=1$.
The moduli action (\ref{saf1}) (with $\bar a=0$) reduces simply to
\begin{equation}
\label{smk1}
S_{\rm moduli}^{(k=1)} =  - 2 \bar\varepsilon\varepsilon \,x^2 -
\frac{\bar\varepsilon}{2}\,
\theta^{\alpha A}\epsilon_{AB}(\tau_3)_{\alpha\beta} \theta^{\beta B}
+ \widehat S_{\rm moduli}^{(k=1)}~~,
\end{equation}
where the last term, containing only the centered moduli $\widehat{\mathcal{M}}^{(k=1)}$
$=\{w,\mu,$ $\chi,D_c,\lambda\}$, is
\begin{equation}
\widehat S_{\rm moduli}^{(k=1)} =
- 2 \bar w_{\dot\alpha} w^{\dot\alpha} \chi^\dagger \left(\chi + a\right) +
\ii\frac{\sqrt{2}}{2}
\chi^\dagger\mu^A \epsilon_{AB}\mu^B
-\ii D_c W^c - \ii \lambda^{\dot\alpha}_{~A}\left(w_{\dot\alpha}\bar \mu^A +
\mu^A \bar w_{\dot\alpha}\right)
\end{equation}
and does not depend on the deformation parameters $\varepsilon$ and $\bar\varepsilon$
\footnote{Notice that all the D(--1)/D(--1) moduli, which in general are $k\times k$ matrices,
reduce just to numbers for $k=1$, and that $a'_\mu$ and $M^{\alpha A}$
contain only their components along the identity, namely the center coordinate
$x_\mu$, and its super-partner $\theta^{\alpha A}$ (cf.
\eq{xtheta}).}.
The corresponding partition function is therefore given by
\begin{equation}
\label{zk1def}
Z^{(k=1)}(a,\varepsilon) =  \int d^4x\, d^4\theta
\,\,\ee^{-2\bar\varepsilon\varepsilon
x^2 - \frac{1}{2}\,\bar\varepsilon\,\theta\cdot\theta}
\,\mathcal{F}^{(k=1)}(a)
= \frac{1}{\varepsilon^2} \,\mathcal{F}^{(k=1)}(a)~~.
\end{equation}
In the last step we have trivially performed the gaussian integration
over $x$ and $\theta$ and produced a factor of $1/\varepsilon^2$, since
the centered partition function
\begin{equation}
\label{fk1a}
\mathcal{F}^{(k=1)}(a) =
\int d\widehat{\mathcal{M}}^{(k=1)} \ee^{-\frac{8\pi^2}{g^2}\,-\,\hat S_{\rm moduli}^{(k=1)}}
\end{equation}
is $\varepsilon,\bar\varepsilon$-independent.
Effectively, in the presence of the full deformation we have the rule
\begin{equation}
\label{rulext}
\int d^4x\, d^4\theta \to \frac{1}{\varepsilon^2}~~.
\end{equation}
It is interesting to remark that the same effective rule appears also in other contexts
related to topological string amplitudes, like in their relation to black
hole free energy recently proposed by Ooguri-Strominger-Vafa
(see in particular Section 3.2 of Ref. \cite{Ooguri:2004zv}).

Turning on the deformations, we can therefore compute the
full partition function $Z(a;\varepsilon)$ by integrating over all the
moduli and convert the would-be (super)-volume
divergences into $\varepsilon$-singularities.
In this respect,
a further important point has to be taken into account.
The combined effect of the scalar v.e.v.'s $a_u$ and of the
$\varepsilon,\bar\varepsilon$-deformations localizes completely
the integral over the moduli space, in the sense that only
point-like solutions contribute. Thus, a trivial superposition of two
instantons of charge $k_1$ and $k_2$
contributes to the sector of charge $k_1 + k_2$. This cluster
decomposition implies \cite{Nekrasov:2002qd,Nekrasov:2005wp} that the localized
integral over the deformed instanton moduli space
of a fixed charge $k$ contains both {\em connected} and
{\em disconnected} contributions.
For $\bar\varepsilon=0$ the disconnected
configurations, consisting of separated instantons of charges
$k_i$ such that $\sum_i k_i = k$, do not contribute
to the sector of charge $k$, but instead they do in the fully localized case
when $\bar\varepsilon\not=0$.
Therefore, the partition function computed via the
localization techniques
corresponds to the exponential of the non-perturbative
prepotential, namely
\begin{equation}
\label{exprel}
\begin{aligned}
Z(a;\varepsilon) &
= \exp\left(\frac{\mathcal{F}_{\mathrm{n.p.}}(a;\varepsilon)}{\varepsilon^2}\right)
= \exp\left(\sum_{k=1}^\infty \frac{\mathcal{F}^{(k)}(a;\varepsilon)}{\varepsilon^2}\right)
\\
& = \exp\left(\sum_{h=0}^\infty \sum_{k=1}^\infty c_{k,h} \frac{\varepsilon^{2h-2}}{a^{2h}}
\left(\frac{\Lambda}{a}\right)^{4k}
\right)
\end{aligned}
\end{equation}
where $\mathcal{F}_{\mathrm{n.p.}}(a;\varepsilon)$ and $\mathcal{F}^{(k)}(a;\varepsilon)$
are the expressions given in Eqs. (\ref{Fnp}), (\ref{cipfw}) and (\ref{Fkpw}) evaluated
for constant values of the scalar and Weyl multiplets.
Notice that the factor of $1/\varepsilon^2$ in the exponent
of \eq{exprel} effectively represents the (super)-volume integral
according to \eq{rulext}, while the disconnected contributions now
cancel.

The relation (\ref{exprel}) allows to check successfully \cite{Klemm:2002pa}
the expression of the
coefficients $c_{k,h}$ obtained from the multi-instanton deformed calculus against
the results from topological string amplitudes, as
originally conjectured in Ref. \cite{Nekrasov:2002qd}.

\vskip 0.8cm
\section{Conclusions}
\label{secn:concl}

Realizing supersymmetric gauge theories by means of fractional D3 branes
allows to compute instanton effects by considering the inclusion of D(--1) branes
and offers a natural way to study the effect of turning on closed string ``gravitational''
backgrounds. In this paper we have considered, in particular, the effect
of including  a self-dual graviphoton field-strength coming from the R-R closed string sector
in a $\mathcal{N}=2$ gauge theory.

We have shown that a constant graviphoton field-strength proportional to $\varepsilon$
exactly produces those
modifications of the instanton sectors which have been advocated in the literature to
fully localize the integration over the moduli. This localization allows to perform
explicitly calculation of the
instanton partition functions $Z_k(a,\varepsilon)$, where
$a$ is the scalar v.e.v., for arbitrary value of the topological charge $k$.
Moreover, we have shown that extending the computation to a dynamical graviphoton
determines a prepotential for the resulting $\mathcal{N}=2$ low-energy effective theory
which includes gravitational F-terms.

These F-terms can be alternatively computed in a different setting, where the low-energy
$\mathcal{N}=2$ effective action is engineered by considering closed strings on a suitable
CY manifold; in this case such couplings are encoded in topological string amplitudes on the
same manifold.
The two different roads to determine these $F$-couplings must lead to the same result.
This is a very natural way to state the conjecture by N. Nekrasov \cite{Nekrasov:2002qd}
that the coefficients arising in the $\varepsilon$-expansion of multi-instanton partition
functions match those appearing in higher genus topological string amplitudes
on CY manifolds.

It would be very nice%
\footnote{We thank N. Nekrasov and M. Vonk for having pointed out to us this issue.}
to be able to follow the fate of the constant RR background that we turn on
in the fractional brane set-up through a series of geometrical operations (including the blow-up of the orbifold)
and string dualities connecting this setup to the local CY set-up.
This does not seem to be a completely trivial task
and this point deserves further investigation.

\vskip 1cm
\section*{Acknowledgements}
We thank C. Bachas, B. de Wit, R. Flume, P.A. Grassi, J.F. Morales Morera, N. Nekrasov, R. Poghossian,
I. Pesando, A. Tanzini and M. Vonk for
several useful discussions.

\noindent
This work is partially supported by the European Community's Human Potential
Programme under contracts MRTN-CT-2004-005104 (in which A.L. is
associated to Torino University), MRTN-CT-2004-503369 and MRTN-CT-2004-512194,
by the INTAS contract 03-51-6346, by the NATO contract NATO-PST-CLG.978785,
and by the Italian MIUR under contracts PRIN-2005023102 and PRIN-2005024045.

\vskip 0.8cm

\appendix

\section{Appendix}
\label{secn:appA}
\setcounter{equation}{0}
\paragraph{$\mathbb{Z}_2$ orbifold:}
Our notation and conventions are as follows: we label the four longitudinal directions of
the D3 branes with indices $\mu,\nu,...=1,2,3,4$, and the six
transverse directions with indices $a,b,...=5,...,10$.
On the complexified internal string coordinates
\begin{equation}
\label{frac1}
\begin{aligned}
Z &\equiv& \frac{X^5 +\ii X^6}{\sqrt{2}}~~~,~~~Z^1 &\equiv& \frac{X^7 +\ii X^8}{\sqrt{2}}
~~~,~~~Z^2 &\equiv& \frac{X^9 +\ii X^{10}}{\sqrt{2}}~~,
\\
\Psi &\equiv& \!\!\!\!\frac{\psi^5 +\ii \psi^6}{\sqrt{2}}~~~,~~~\Psi^1
&\equiv&
\frac{\psi^7 +\ii \psi^8}{\sqrt{2}}
~~~,~~~\Psi^2 &\equiv& \frac{\psi^9 +\ii \psi^{10}}{\sqrt{2}}~~,
\end{aligned}
\end{equation}
the $\mathbb{Z}_2$ orbifold generator $h$ acts as
\begin{equation}
\label{frac2}
h~:~
\left\{\begin{array}{c}
(Z,Z^1,Z^2) ~\rightarrow~ (Z,-Z^1,-Z^2)
\\
(\Psi,\Psi^1,\Psi^2) ~\rightarrow~ (\Psi,-\Psi^1,-\Psi^2)
\end{array}\right.
~~.
\end{equation}
Under the $\mathrm{SO}(10)\rightarrow
\mathrm{SO(4)}\times \mathrm{SO(6)}$ decomposition induced by the
presence of the D3 branes, the ten-dimensional (anti-chiral) spin
fields $S^{\dot{\mathcal{A}}}$ ($\dot{\mathcal{A}}=1,\ldots,16$)
of the RNS formalism become products of four- and six-dimensional spin
fields according to
\begin{equation}
\label{spindec}
S^{\dot{\mathcal{A}}} ~\to~ (S_{\alpha}S_{A'}\,,\,S^{\dot\alpha}S^{A'})
\end{equation}
where the index $\alpha$ (or $\dot\alpha$) denotes positive (or negative)
chirality in four dimensions, and the upper (or lower) index $A'$
labels the chiral (or anti-chiral) spinor representation of
$\mathrm{SO}(6)$. Under the further
$\mathrm{SO(6)}\rightarrow \mathrm{SO(2)}\times \mathrm{SO(4)}$
breaking induced by the orbifold projection, a chiral
$\mathrm{SO}(6)$ spinor $S^{A'}$
splits into $\big(+\frac{1}{2};(\mathbf{2},\mathbf{1})\big)
+ \big(-\frac{1}{2};(\mathbf{1},\mathbf{2})\big)$,
labeled respectively by an upper index $A=1,2$ and $\hat A=3,4$,
while an anti-chiral $\mathrm{SO}(6)$ spinor $S_{A'}$ splits into
 $\big(+\frac{1}{2};(\mathbf{1},\mathbf{2})\big)
+ \big(-\frac{1}{2};(\mathbf{2},\mathbf{1})\big)$,
labeled respectively by a lower index $A=1,2$ and $\hat A=3,4$.

On the internal spinor indices the orbifold generator acts as
a $\mathrm{SO}(4)$ chirality operator
\footnote{The indices $\pm$ appearing in
the table (\ref{spintransf}) denote the charge $\pm1/2$ carried by the spin field
under the three bosons that bosonize the three world-sheet spinors
$\Psi$, $\Psi^{1}$ and $\Psi^{2}$.} as follows
\begin{equation}
\label{spintransf}
\begin{tabular}{c|c|c}
$S_{A'}$ & $S^{A'}$   & $h$  \\
\hline
$\Big.S_{1}=S_{---}$ & $S^{1}=S^{+++}$ & $+1$   \\
$\Big.S_{2}=S_{-++}$ & $S^{2}=S^{+--}$ & $+1$   \\
$\Big.S_{3}=S_{+-+}$ & $S^{3}=S^{-+-}$ & $-1$   \\
$\Big.S_{4}=S_{++-}$ & $S^{4}=S^{--+}$ & $-1$
\end{tabular}
\end{equation}
Thus, the spinors belonging to the
$\big(\pm \frac{1}{2};(\mathbf{2},\mathbf{1})\big)$ representation
are even under the orbifold projection while the ones transforming in
the $\big(\pm\frac{1}{2};(\mathbf{1},\mathbf{2})\big)$ representation are odd.

\paragraph{$\mathbf{d=4}$ Clifford algebra:}

The matrices $(\sigma^\mu)_{\alpha\dot\beta}$ and
$(\bar\sigma^{\mu})^{\dot\alpha\beta}$ which generate the
Clifford algebra in four dimensions are defined as
\begin{equation}
\label{sigmas}
\sigma^\mu =
(\uno,-\ii\vec\tau)~,\hskip 0.8cm
\bar\sigma^\mu =
\sigma_\mu^\dagger = (\uno,\ii\vec\tau)~~,
\end{equation}
where $\tau^c$ are the ordinary Pauli matrices.

Out of these matrices, the $\mathrm{SO}(4)$ generators are defined by
\begin{equation}
\label{sigmamunu}
\sigma_{\mu\nu}
=\frac 12(\sigma_\mu\bar\sigma_\nu -
\sigma_\nu\bar\sigma_\mu)~,
\hskip 0.8cm
\bar\sigma_{\mu\nu}
=\frac 12(\bar\sigma_\mu\sigma_\nu -
\bar\sigma_\nu\sigma_\mu)~~;
\end{equation}
the matrices $\sigma_{\mu\nu}$ are self-dual and thus generate the
$\mathrm{SU}(2)_\mathrm{L}$ factor of $\mathrm{SO}(4)$;
the anti-self-dual matrices $\bar\sigma_{\mu\nu}$ generate
instead the $\mathrm{SU}(2)_\mathrm{R}$ factor.
The explicit mapping of a self-dual $\mathrm{SO}(4)$
tensor into the adjoint representation of the $\mathrm{SU}(2)_\mathrm{L}$ factor
is realized by the 't Hooft symbols $\eta^c_{\mu\nu}$; the analogous
mapping of an anti-self dual tensor into the adjoint
of the $\mathrm{SU}(2)_\mathrm{R}$ subgroup is realized by
$\bar\eta^c_{\mu\nu}$. One has
\begin{equation}
\label{sigmaeta}
(\sigma_{\mu\nu})_{\alpha}^{~\beta} =
\ii \,\eta^c_{\mu\nu}\, (\tau^c)_{\alpha}^{~\beta}
,\qquad
(\bar\sigma_{\mu\nu})^{\dot\alpha}_{~\dot\beta} = \ii \,\bar\eta^c_{\mu\nu}\,
(\tau^c)^{\dot\alpha}_{~\dot\beta}~~.
\end{equation}

\paragraph{String field correlators:}

The non-trivial OPE's of the world-sheet fields with space-time indices that are
used in the main text are
\begin{equation}
\label{stcorr}
\begin{aligned}
&S^{\dot\alpha}(z) \,S_\beta(w)\sim
\frac{1}{\sqrt{2}}\,
(\bar \sigma^\mu)^{\dot\alpha}_{~\beta}\, \psi_\mu(w)&
~~~~,~~~~
&S^{\dot\alpha}(z)\, S^{\dot\beta}(w) \sim
- \,\frac{\epsilon^{\dot\alpha\,\dot\beta}}{(z - w)^{1/2}}&
\\
&S_{\alpha}(z) \,S_{\beta}(w) \sim
\frac{\epsilon_{\alpha\beta}}{(z - w)^{1/2}}&
~~~~,~~~~
&\psi^\mu(z) \,S^{\dot\alpha}(w) \sim
\frac{1}{\sqrt{2}}\,
\frac{(\bar \sigma^\mu)^{\dot\alpha\beta}\, S_\beta(w)}{(z -
w)^{1/2}}&
\end{aligned}
\end{equation}
Other relations can be obtained from Eq. (\ref{stcorr}) by
suitable changes of chirality.

The relevant OPE's between fields with internal indices are instead
\begin{equation}
\label{icorr}
\begin{aligned}
&S^A(z)\, S_B(w) \sim~
\frac{\ii \,\delta^A_{~B}}{(z - w)^{3/4}}~~,&~~~~
&S^A(z) \,S^B(w)
\sim~
-\,
\frac{\ii\,\epsilon^{AB}\Psi(w)}{(z - w)^{1/4}}~~,&
\\
&S^{\hat A}(z) \,S^{\hat B}(w) \sim~
-\,
\frac{\ii\,\epsilon^{{\hat A}{\hat B}}\bar\Psi(w)}{(z - w)^{1/4}}~~,&~~~~
& S_A(z) \,S_B(w) \sim~
\frac{\ii\,\epsilon_{AB}\bar \Psi(w)}{(z - w)^{1/4}}
~~,&
\\
&S_{\hat A}(z) \,S_{\hat B}(w) \sim~
\frac{\ii\,\epsilon_{{\hat A}{\hat B}}\Psi(w)}{(z - w)^{1/4}}~~,&~~~~
&\Psi(z)\, S_A(w) \sim~
\frac{\epsilon_{AB} S^B(w)}{(z - w)^{1/2}}
~~,&
\\
&{\bar\Psi}(z)\, S_{\hat A}(w) \sim~
\frac{\epsilon_{{\hat A}{\hat B}} S^{\hat B}(w)}{(z -
w)^{1/2}}~~,&~~~~
&\Psi(z)\, S^{\hat A}(w) \sim~
\frac{\epsilon_{{\hat A}{\hat B}}S^{\hat B}(w)}{(z - w)^{1/2}}
~~,&
\\
&{\bar\Psi}(z)\, S^A(w) \sim~
\frac{\epsilon_{AB} S^{B}(w)}{(z - w)^{1/2}}~~,&~~~~
&\Psi(z)\,\bar\Psi(w) \sim~ \frac{1}{z-w}~~.&
\end{aligned}
\end{equation}
{F}rom these OPE's we can derive the following 3- and 4-point correlators which
are needed for the calculation of the scattering amplitudes
presented in the main text
\begin{equation}
\label{3corr}
\begin{aligned}
&\big\langle
\Psi(z_1)\,S_A(z_2)\,S_B(z_3)\big\rangle
=\frac{\ii \,\epsilon_{AB}}{(z_1-z_2)^{1/2}
(z_1-z_3)^{-1/2}(z_2-z_3)^{1/4}}~~,&
\\
&\big\langle
\bar\Psi(z_1)\,S_{\hat A}(z_2)\,S_{\hat B}(z_3)\big\rangle
=
\frac{\ii\, \epsilon_{{\hat A}{\hat B}}}{(z_1-z_2)^{1/2}
(z_1-z_3)^{1/2}(z_2-z_3)^{1/4}}~~,&
\end{aligned}
\end{equation}
and
\begin{equation}
\big\langle \psi^{\mu}(z_1) \, \psi^{\nu}(z_2)\, S_{\alpha}(w) \,
S_{\beta}(\bar w) \big\rangle =
A\,\delta^{\mu\nu}\,\epsilon_{\alpha\beta} +
B (\sigma^{\mu\nu})_{\alpha\beta}~~,
\label{4point}
\end{equation}
where
\begin{equation}
\label{A}
A= \frac{1}{2}\,
\frac{(z_1-w)(z_2-\bar w)+(z_2-w)(z_1-\bar w)}{(z_1-z_2)
\big[(z_1-w)(z_1-\bar w)(z_2-w)(z_2-\bar w)(w-\bar w)\big]^{{1}/{2}}}
\end{equation}
and
\begin{equation}
\label{B}
B = -\frac{1}{2}\,
\frac{(w-\bar w)^{{1}/{2}}}{\big[(z_1-w)(z_1-\bar w)(z_2-w)(z_2-\bar w)\big]^{{1}/{2}}}
~~.
\end{equation}

\paragraph{Bosonic twist fields:}
For the open strings that stretch between a D3 and a D(--1) brane, the string fields $X^\mu$
along the D3 brane world-volume have mixed Neumann-Dirichlet
boundary conditions, which can be seen as due to
twist and anti-twist fields $\Delta(z)$ and
$\bar\Delta(z)$. These fields change the boundary conditions from Neumann to
Dirichlet and vice-versa by introducing a cut in the world-sheet
(see for example Ref.~\cite{Dixon:1985}).
The twist fields $\Delta(z)$ and $\bar\Delta(z)$ are bosonic
operators with conformal dimension $1/4$ and their OPE's are
\begin{equation}
\Delta(z_1)\,\bar\Delta(z_2) \sim (z_1-z_2)^{1/2}~~~, ~~~
\bar\Delta(z_1)\,\Delta(z_2) \sim -\,(z_1-z_2)^{1/2} ~~,
\label{deltadelta}
\end{equation}
where the minus sign in the second correlator is  an ``effective''
rule to correctly account for the space-time statistics in correlation
functions.

\paragraph{Superghost correlators:}
\begin{equation}
\label{3sghostcorr}
\langle\ee^{-\varphi(z_1)}\,\ee^{-\frac{1}{2}\varphi(z_2)}\,
\ee^{-\frac{1}{2}\varphi(z_3)}\rangle
= (z_1-z_2)^{-{1}/{2}}\,(z_1-z_3)^{-{1}/{2}}\,
(z_2-z_3)^{-{1}/{4}}~~,
\end{equation}
\begin{equation}
\label{4sghostcorr}
\begin{aligned}
{}&\!\!\!\!
\langle\ee^{-\frac{1}{2}\varphi(z_1)}\,\ee^{-\frac{1}{2}\varphi(z_2)}\,
\ee^{-\frac{1}{2}\varphi(z_3)}\,
\ee^{-\frac{1}{2}\varphi(z_4)}\rangle
\\
{}&~~= \Big[(z_1-z_2)\,(z_1-z_3)\,(z_1-z_4)\,(z_2 - z_3)\,
(z_2 - z_4)\,(z_3 - z_4)\Big]^{-{1}/{4}}~~.
\end{aligned}
\end{equation}

\end{document}